\documentclass[12pt]{article}
\usepackage{amsmath}
\usepackage{graphicx}
\usepackage{enumerate}
\usepackage{natbib}
\usepackage{amsmath}
\usepackage{url} 

\newcommand{\blind}{0}

\DeclareMathAlphabet{\mybm}{OT1}{ptm}{b}{it}

\usepackage{mathptmx}
\usepackage{euscript}
\usepackage{latexsym}
\usepackage{amsmath}
\usepackage{amsfonts}
\usepackage{amssymb}
\usepackage{ulem}
\usepackage{amsthm}

\newtheorem{thm}{Theorem}

{\theoremstyle{remark} 
}

{\theoremstyle{definition} 

}

\newcommand{\BIBO}{{\small \BIBO}}

\newcommand{\eq}{\begin{eqnarray*}}
\newcommand{\eqq}{\end{eqnarray*}}
\newcommand{\eqn}{\begin{eqnarray}}
\newcommand{\eqqn}{\end{eqnarray}}

\newcommand{\eqb}{\begin{align*}}
\newcommand{\eqqb}{\end{align*}}

\newcommand{\eqna}{\begin{align}}
\newcommand{\eqqna}{\end{align}}

\newcommand{\bA}{\mathbf{A}}

\newcommand{\bD}{\mathbf{D}}

\newcommand{\bI}{\mathbf{I}}

\newcommand{\bX}{\mathbf{X}}
\newcommand{\ba}{\mathbf{a}}
\newcommand{\bb}{\mathbf{b}}
\newcommand{\bc}{\mathbf{c}}

\newcommand{\bx}{\mathbf{x}}
\newcommand{\by}{\mathbf{y}}
\newcommand{\bz}{\mathbf{z}}

\newcommand{\bu}{\mathbf{u}}
\newcommand{\bv}{\mathbf{v}}
\newcommand{\br}{\mathbf{r}}
\newcommand{\bs}{\mathbf{s}}

\newcommand{\bw}{\mathbf{w}}

\newcommand{\bbR}{\mathbb{R}}

\newcommand{\0}{{\bf 0}}
\newcommand{\1}{{\bf 1}}

\newcommand{\cF}{\EuScript{F}}

\newcommand{\ep}{\epsilon}

\newcommand{\om}{\omega}

\newcommand{\bmdel}{\pmb{\delta}}
\newcommand{\bmbeta}{\pmb{\beta}}
\newcommand{\bmgam}{\pmb{\gamma}}

\newcommand{\bmphi}{\pmb{\phi}}

\newcommand{\bmth}{{\pmb{\theta}}}

\newcommand{\bmxi}{\pmb{\xi}}

\newcommand{\bmzeta}{\pmb{\zeta}}

\newcommand{\bmlam}{\pmb{\lambda}}

\newcommand{\bPhi}{\pmb{\Phi}}

\usepackage{float}
\usepackage{caption}
\usepackage{subcaption}
\usepackage{booktabs}

\def\spacingset#1{\renewcommand{\baselinestretch}%
{#1}\small\normalsize} \spacingset{1}

 \DeclareMathSymbol{,}{\mathpunct}{operators}{"2C}

\begin{document}

\if0\blind
{
  \title{\bf  Spline Quantile Regression}
  \author{Ta-Hsin Li\footnote{Formerly affiliated with IBM T.\ J.\ Watson Research Center. 
  Email: {\sc thl024@outlook.com}} \ and Nimrod Megiddo\footnote{IBM T.\ J.\ Watson
  Research Center, Yorktown Hights, NY 10598. Email: {\sc megiddo@us.ibm.com}}
  }
  \date{April 8, 2025}
  \maketitle
} \fi

\if1\blind
{
  \title{\bf   Spline Quantile Regression} 
  \maketitle
} \fi

\begin{abstract}
Quantile regression is a powerful tool capable of offering a richer view of the data as compared to 
least-squares regression.  Quantile regression is typically performed individually
on a few quantiles or a grid of quantiles without considering the similarity of 
the underlying regression coefficients at nearby quantiles. When needed, an ad hoc post-processing 
procedure such as kernel smoothing is employed to smooth the individually estimated coefficients across quantiles
and thereby  improve the performance of these estimates. This paper introduces a new method, called
spline quantile regression (SQR), that unifies quantile regression with quantile smoothing and jointly 
estimates the regression coefficients across quantiles as smoothing splines. We discuss the computation
of the SQR solution as a linear program (LP) using an interior-point algorithm. 
We also experiment with some gradient algorithms that require less memory than the LP algorithm.
The performance of the SQR method and these algorithms is evaluated using simulated and real-world data.

\bigskip\bigskip
\noindent
{\it Keywords}: function estimation, gradient descent, interior point, linear program, 
quantile periodogram, quantile regression, smoothing, spline 

\bigskip\bigskip
\noindent
{\it Acknowledgment}: The authors would like to thank Dr.\ R.\ Koenker for advice on 
the FORTRAN code of the interior-point algorithm.
\vfill
\end{abstract}

\newpage
\spacingset{1.9} 

\section{Introduction}

Quantile regression  (QR) is a powerful statistical tool that complements the conventional 
least-squares regression (Koenker and Bassett 1978; Koenker 2005). Quantile regression postulates 
the conditional quantile of a dependent variable, rather than the conditional mean,  
as a function of the explanatory variables. By varying the quantile level, one can 
explore this relationship across all quantiles and thereby obtains  a richer view 
of the data than that offered by least-squares regression.
Recent years have witnessed further development 
of the quantile regression method, Examples include, just to name a few,  the fast algorithms to compute quantile regression when the number of regressors is very large (He et al.\ 2023), the statistical analyses for the regression coefficients as functions of the quantile level (Belloni et al.\ 2019; Hao et al.\ 2023), and the
techniques to overcome quantile crossings for applications
that require the quantiles to be monotone at any point in the space of regressors (He 1997; 
Wu and Liu 2009; Bondell et al.\ 2010).

In this paper, we are interested in situations where the underlying quantile regression coefficients 
vary smoothly across quantiles. In typical applications, the smoothness
is either ignored entirely by performing quantile regression  independently at different quantiles,
or handled separately by employing an ad hoc post-processing procedure such as kernel smoothing
to smooth the raw estimates across quantiles (Koenker 2005, pp.\ 158--159).

We offer an alternative solution, called spline quantile regression (SQR). 
The SQR method extends the original QR problem  at an individual quantile 
into a function estimation problem across all quatiles. In this problem, the regression coefficients are treated as smooth functions of the quantile level in a space of splines, and the resulting model is fitted to the data jointly on a grid of quantiles.
A penalty term is added to the QR cost function to regularize the smoothness of the functional regression
coefficients. The SQR method unifies quantile regression with quantile smoothing and provides an 
estimator of regression coefficients as  smoothing splines. In this paper, we focus on the computation
of the SQR solution. We evaluate and demonstrate the SQR solution with simulated and real data.

Smoothing splines have been used in the context of quantile regression to represent nonparametric regression functions (Koenker et al.\ 1994; Oh et al.\ 2011; Andriyana et al.\ 2014; He et al.\ 2021). In these methods, the smoothness of the regression function is regularized  in a way similar to the spline smoothing problem under the least-squares framework (Wahba 1975). The SQR problem, on the other hand, represents 
the regression coefficients  as spline functions of the quantile level. The resulting model is a linear function of the regressors for  fixed quantile level, and a nonlinear smooth function of the quantile level
for fixed  regressors.

To retain the numerical characteristics of quantile regression, we employ the integral of the $L_1$-norm
of second derivatives as the penalty to regularize the functional regression coefficients. 
A similar measure was employed in  Koenker et al.\ (1994) for nonparametric
quantile regression. We show  that the resulting SQR problem can be reformulated as a 
primal-dual pair of linear program (LP) and solved using an interior-point algorithm 
developed by Portnoy and Koenker (1997) for the ordinary QR  problem.

In this paper, we also experiment with some gradient algorithms as computationally 
more efficient  alternatives to approximate the LP solution. This investigation is largely motivated 
by the success of gradient algorithms in training neural network models 
with non-smooth objective functions (Goodfellow et al.\ 2016; Ruder 2016).
The success stories have generated renewed interest in trying to better understand 
the behavior of gradient algorithms  in non-smooth situations (e.g., Lewis and Overton 2013;
Asl and Overton 2020). As in machine learning applications, our aim is not to use these algorithms 
as a replacement of LP to produce the exact solution. Instead,  we are interested in algorithms 
that provide sufficiently good approximations to the LP solution but with reduced computational burden, 
especially the demand on computer memory. 

The remainder is organized as follows. Section 2 describes the SQR problem. Section 3 discusses the LP reformulation and proves the suitability of the interior-point algorithm. Section 4 presents some numerical examples with simulated and real-word data to demonstrate the SQR method. Section 5 describes
the BFGS, ADAM, and GRAD algorithms. Section 6 contains the experimental results of gradient algorithms. Concluding remarks are given in Section 7. 
The R functions that implement the SQR method are described in Appendix.

\section{Spline Quantile Regression}

Let $\{ y_t: t=1,\dots,n\}$ be a sequence of $n$ observations of a dependent variable and $\{ \bx_t: t=1,\dots,n\}$ be the corresponding values of a $p$-dimensional regressor.  Under the quantile regression framework (Koenker 2005), it is assumed that the conditional quantile of $y_t$ at a quantile level $\tau \in (0,1)$ given $\bx_t$ can be expressed as
\eqn
Q_{y_t}(\tau \mid \bx_t) = \bx_t^T \bmbeta(\tau).
\label{qr}
\eqqn
Given an increasing sequence of quantile levels $\{ \tau_\ell: \ell =1,\dots,L\} \subset (0,1)$, the standard 
QR method produces an estimate for each $\bmbeta_\ell := \bmbeta(\tau_\ell)$ independently by solving
\eqn
\hat{\bmbeta}_\ell
:=  \operatorname*{argmin}_{\bmbeta  \in \bbR^p} 
 \sum_{t=1}^n
\rho_{\tau_\ell}(y_t - \bx_t^T \bmbeta) \quad (\ell=1,\dots,L),
\label{qr}
\eqqn
where $\rho_\tau(y) := y (\tau - {\cal I}(y < 0))$ denotes the objective function of quantile regression 
at quantile level $\tau$ (Koenker 2005, p.\ 5). In typical situations, the quantile levels in $\{ \tau_\ell\}$
are specified  by the user based on the need of the application. In rare occasions, one may want to consider
the quantiles that correspond to all distinct  quantile regression solutions (Portnoy 1991; Koenker 2005, p.\ 303)
similar to the quantiles associated with the order statistics.

In this paper, we are interested in estimating $\bmbeta(\tau)$ in (\ref{qr}) as a function of $\tau$
in a closed subinterval of $(0,1)$, e.g., $[\ep,1-\ep] \subset (0,1)$ for some $\ep \in (0,0.5)$.
We propose to obtain an estimate by solving the spline quantile regression (SQR) problem
\eqn
\hat{\bmbeta}(\cdot) := \operatorname*{argmin}_{\bmbeta(\cdot) \in \cF^p} \bigg\{
n^{-1}
\sum_{\ell=1}^L \sum_{t=1}^n
\rho_{\tau_\ell}(y_t - \bx_t^T \bmbeta(\tau_\ell)) +  c \sum_{\ell=1}^L w_\ell \,  \|\ddot{\bmbeta}(\tau_\ell)\|_1 \bigg\},
\label{sqr}
\eqqn
where  $\cF$ is a functional space spanned by spline basis functions, $c \ge 0$  is a smoothing 
or penalty parameter, and $\{ w_\ell \}$ is a user-specified sequence of nonnegative constants that allow different 
contributions from different quantiles to the penalty term. 

We employ the $L_1$-norm of second derivatives  in (\ref{sqr}) 
as the roughness measure of the functional coefficient $\bmbeta(\cdot)$ in order to retain 
the LP characteristics of the original QR  problem  (Koenker 2005). 
This is the same reason offered in Koenker et al.\ (1994) to justify the $L_1$-norm penalty
for the spline-based nonparametric regressor in a nonparamtric quantile regression problem.
It leads to a piecewise linear function that can be used to approximate any continuous functions
when $\max\{ |\tau_{\ell+1}-\tau_\ell|\}$ is sufficiently small.
An altervative to the $L_1$-norm is the squared $L_2$-norm, which is commonly used 
for spline smoothing under the least-squares framework (Wahba 1975).  This option would result in a quadratic program (QP) instead of an LP.  We choose to deal with this  problem  elsewhere.

Let $\{ \phi_k(\tau): k=1,\dots,K\}$ denote a set of spline basis functions of $\tau \in (0,1)$. 
Then, any function $\beta_j(\tau)$ in $\cF$ can be expressed as
\eq
\beta_j(\tau) = \sum_{k=1}^K \phi_k(\tau) \theta_{jk} =  \bmphi^T(\tau) \bmth_j \quad (j=1,\dots,p),
\eqq
where
\eq
\bmphi(\tau)  :=  [\phi_1(\tau),\dots,\phi_K(\tau)]^T \in \bbR^K, \quad 
\bmth_j \ :=    [\theta_{j1},\dots,\theta_{jK}]^T \in \bbR^{K}.
\eqq
Therefore, for any $\bmbeta(\tau) := [\beta_1(\tau),\dots, \beta_p(\tau)]^T \in \cF^p$, we can write
\eq
\bmbeta(\tau)  = \bPhi(\tau) \bmth,
\eqq
where
\eq
\bPhi(\tau) :=  \bI_p \otimes \bmphi^T(\tau) \in \bbR^{p \times pK}, \quad \bmth  :=  [\bmth_{1}^T,\dots,\bmth_{p}^T]^T \in \bbR^{pK}.
\eqq
With this notation and $c_\ell := n c w_\ell$, the SQR problem (\ref{sqr}) can be restated as
\eqn
\hat{\bmth} :=
\operatorname*{argmin}_{\bmth \in \bbR^{pK}} \bigg\{
 \sum_{\ell=1}^L \sum_{t=1}^n
\rho_{\tau_\ell}(y_t - \bx_t^T \bPhi(\tau_\ell) \bmth) +  \sum_{\ell=1}^L  c_\ell \,
 \| \ddot{\bPhi}(\tau_\ell) \bmth \|_1 \bigg\}
\label{sqr2}
\eqqn
and
\eqn
\hat{\bmbeta}(\tau) := \bPhi(\tau) \hat{\bmth}.
\label{betah}
\eqqn
In other words, the SQR problem (\ref{sqr}) can be solved by searching for the vector $\hat{\bmth}$ in $\bbR^{pK}$ according to (\ref{sqr2}) and then converting it to the desired function $\hat{\bmbeta}(\cdot)$ according to (\ref{betah}).

The SQR problem in (\ref{sqr})  is different from the problem 
of quantile smoothing splines considered by Koenker et al.\ (1994).
The latter uses splines to represent nonparametric functions of independent variables in form
of $Q_y(\tau \mid x) = s(x)$ for fixed $\tau$.
The SQR problem in (\ref{sqr}) is also different from the problems considered by Andriyana et al.\ (2014) and Kim (2007)
in which splines are used to represent regression coefficients as functions of an auxiliary variable rather than
functions of the quantile level $\tau$.

\section{Linear Program Reformulation}

Like the ordinary QR  problem (Koenker 2005, p.\ 7), 
the SQR problem in (\ref{sqr2}) can be reformulated as a linear program (LP) with nonnegative decision variables:
\eqn
\lefteqn{
(\hat{\bmgam}, \hat{\bmdel}, \hat{\bu}_1, \hat{\bv}_1, \hat{\br}_1, \hat{\bs}_1,\dots,
\hat{\bu}_L, \hat{\bv}_L, \hat{\br}_L, \hat{\bs}_L )  :=} \notag \\
&  &  
\operatorname*{argmin}_{
 (\bmgam,\bmdel,\bu_1,\bv_1,\br_1,\bs_1,\dots,\bu_L,\bv_L,\br_L,\bs_L) \in \bbR_+^{d}} \
\sum_{\ell=1}^L \{ \tau_\ell \1_n^T \bu_\ell + (1-\tau_\ell) \1_n^T \bv_\ell + \1_p^T \br_\ell + \1_p^T \bs_\ell \}
   \notag \\
&& \quad
{\rm s.t.} \quad
\left\{
\begin{array}{l} 
\bX \, \bPhi(\tau_\ell) (\bmgam - \bmdel) + \bu_\ell - \bv_\ell = \by \quad (\ell=1,\dots,L),  \\
c_\ell \, \ddot{\bPhi}(\tau_\ell)  (\bmgam - \bmdel) - (\br_\ell - \bs_\ell) = \0\quad (\ell=1,\dots,L), 
\end{array} \right.
\label{lp}
\eqqn
where $\1_n$ and $\1_p$ are the $n$-dimensional and $p$-dimensional vectors of 1's, 
$\bX  :=  [  \bx_1,\dots,\bx_n]^T$ is the regression design matrix, and 
$d := 2pK + 2nL+2pL$ is the total number of decision variables to be optimized.
Among the decision variables in (\ref{lp}), $\bmgam \in \bbR_+^{pK}$ 
and $\bmdel \in \bbR_+^{pK}$ are primary variables which determine the solution $\hat{\bmth}$ in  (\ref{sqr2}) such that
\eqn
\hat{\bmth} = \hat{\bmgam} - \hat{\bmdel}.
\label{theta}
\eqqn
The remaining variables $\bu_\ell \in \bbR_+^n$, $\bv_\ell \in \bbR_+^n$, $\br_\ell \in \bbR_+^p$, and $\bs_\ell \in \bbR_+^p$ are auxiliary variables introduced  just for the purpose of linearizing the objective function in  (\ref{sqr2}).

In the canonical form, the LP problem (\ref{lp}) can be expressed as
\eqn
\min\{ \bc^T \bmxi | \bA \bmxi = \bb; \bmxi \in \bbR_+^{2pK + 2nL+2pL} \},
\label{primal}
\eqqn
where 
\eq
\bc := [\0_p^T,\0_p^T,  \tau_1 \1_n^T, (1-\tau_1) \1_n^T, \1_p^T, \1_p^T,
\dots, \tau_L \1_n^T, (1-\tau_L) \1_n^T, \1_p^T, \1_p^T]^T,
\eqq
\eq
\bmxi := [\bmgam^T,\bmdel^T,\bu_1^T,\bv_1^T,\br_1^T,\bs_1^T,\dots,\bu_L^T,\bv_L^T,\br_L^T,\bs_L^T]^T,
\eqq
\eq
\bA := \left[
\begin{array}{ccccccccccccc}
\bX\, \bPhi(\tau_1)  & - \bX \,\bPhi(\tau_1)  & \bI_n & - \bI_n & \0  & \0  &   \\
\vdots & \vdots   &         &             &         &          & \ddots  \\
\bX\,\bPhi(\tau_L)  & - \bX \,\bPhi(\tau_L)  &        &              &         &         &   &  \bI_n & - \bI_n & \0  & \0     \\
c_1 \, \ddot{\bPhi}(\tau_1) & - c_1 \, \ddot{\bPhi}(\tau_1)  & \0 &  \0 & \bI_p  & \bI_p  &   \\
\vdots & \vdots   &         &             &         &          & \ddots  \\
c_L \, \ddot{\bPhi}(\tau_L) & - c_L \, \ddot{\bPhi}(\tau_L)  &        &              &         &         &   &  \0 & \0 & \bI_p  & -\bI_p   
\end{array}
\right],
\eqq
and
\eq
\bb := [\underbrace{\by^T,\dots,\by^T}_{\text{$L$ times}},\underbrace{\0_p^T,\dots,\0_p^T}_{\text{$L$ times}}]^T \in \bbR^{nL+pL}.
\eqq
Given the canonical form (\ref{primal}), one can compute the SQR solution by any general-purpose LP solvers in open-source or commercial software packages. For example, the {\tt lp} function in the R package `lpSolve'
(Berkelaar 2022)  provides an interface to the open-source software {\tt lp\_solve} that solves LP problems by a simplex method. 
However, in using these solvers, the high dimensionality of decision variables and constraints in (\ref{primal})  can be a challenge to both computer memory and computer time.

In the following, we present a special-purpose algorithm based on the primal-dual 
interior-point method of Portnoy and Koenker (1997). This method  was 
originally employed to solve the ordinary QR problem (Koenker 2005, pp.\ 199--202). 
As it turns out, the SQR problem (\ref{sqr2}) can also be solved by this method with suitable modifications.

First, let us consider the dual LP problem associated with the primal problem (\ref{primal}):
\eqn
\max\{ \bb^T \bmlam | \bA^T \bmlam \le \bc; \bmlam \in \bbR^{nL+pL} \}.
\label{dual0}
\eqqn
The quantity $\bmlam$ may be interpreted as the Lagrange multiplier 
for the equality constraints in (\ref{primal}). 

\begin{thm}
The dual LP problem (\ref{dual0})  can be rewritten as
\eqn
\max\{ \bb^T \bmzeta | \bD^T \bmzeta = \ba;  \bmzeta \in [0,1]^{nL+pL} \},
\label{dual}
\eqqn
where
\eq
\bD &:=& [\bPhi^T(\tau_1) \, \bX^T,\dots,\bPhi^T(\tau_L) \, \bX^T,2 c_1 \, \ddot{\bPhi}^T(\tau_1),\dots,
2 c_L \, \ddot{\bPhi}^T(\tau_L)]^T, \\
\ba & := & \sum_{\ell=1}^L \{ (1-\tau_\ell)  \, \bPhi^T(\tau_\ell) \, \bX^T \1_n + c_\ell \, \ddot{\bPhi}^T(\tau_\ell) \1_p \}.
\eqq
\end{thm}

\noindent
{\it Proof}. By partitioning $\bmlam$ according to the structure of $\bb$
such that 
\eq
\bmlam := [\bmlam_1^T,\dots,\bmlam_L^T,\bmlam_{L+1}^T,\dots,\bmlam_{2L}^T]^T,
\eqq  
the inequalities $\bA^T \bmlam \le \bc$ in (\ref{dual0}) can be written more elaborately as
\eq
\sum_{\ell=1}^L \{ \bPhi^T(\tau_\ell) \, \bX^T \bmlam_\ell + c_\ell \, \ddot{\bPhi}^T(\tau_\ell) \bmlam_{L+\ell} \}  & \le &  \0_p, \\
- \sum_{\ell=1}^L \{ \bPhi^T(\tau_\ell) \, \bX^T \bmlam_\ell + c_\ell \, \ddot{\bPhi}^T(\tau_\ell) \bmlam_{L+\ell}  \} & \le &   \0_p, 
\eqq
\eq
\bmlam_\ell \le \tau_\ell \1_n,   \quad  - \bmlam_\ell \le (1-\tau_\ell) \1_n \quad (\ell=1,\dots,L), \\
 - \bmlam_{L+\ell} \le  \1_p, \quad \bmlam_{L+\ell} \le  \1_p \quad (\ell=1,\dots,L).
\eqq
These inequalities are equivalent to
\eq
\sum_{\ell=1}^L \{\bPhi^T(\tau_\ell) \, \bX^T \bmlam_\ell + c_\ell \, \ddot{\bPhi}^T(\tau_\ell) \bmlam_{L+\ell} \} = \0_p, \\
\bmlam_\ell \in [\tau_\ell-1,\tau_\ell]^n \quad (\ell =1,\dots,L), \\
\bmlam_{L+\ell} \in [-1, 1]^p \quad (\ell =1,\dots,L).
\eqq
By a change of variables,
\eq
\bmzeta_\ell := \bmlam_\ell + (1-\tau_\ell) \1_n, \quad \bmzeta_{L+\ell} :=  \frac{1}{2} (  \bmlam_{L+\ell} + \1_p),
\eqq
we obtain
\eq
\bb^T \bmlam  = \sum_{\ell=1}^L \by^T \bmlam_\ell = \sum_{\ell=1}^L \by^T \bmzeta_\ell  - \sum_{\ell=1}^L (1-\tau_\ell)  
\by^T \1_n = \bb^T \bmzeta + \text{constant},
\eqq
\eq
\lefteqn{\sum_{\ell=1}^L \{\bPhi^T(\tau_\ell) \, \bX^T \bmlam_\ell + c_\ell  \, \ddot{\bPhi}^T(\tau_\ell) \bmlam_{L+\ell} \} } \\
& = & \sum_{\ell=1}^L  \{\bPhi^T(\tau_\ell) \, \bX^T  \bmzeta_\ell + 2 c_\ell \, \ddot{\bPhi}^T(\tau_\ell) \bmzeta_{L+\ell} \} 
 - \sum_{\ell=1}^L \{ (1-\tau_\ell) \, \bPhi^T(\tau_\ell) \, \bX^T \1_n + c_\ell  \, \ddot{\bPhi}^T(\tau_\ell) \1_p \}, 
\eqq
\eq
\bmlam_\ell \in [\tau_\ell-1,\tau_\ell]^n & \leftrightarrow & \bmzeta_\ell \in [0,1]^n, \\
\bmlam_{L+\ell} \in [-1,1]^p & \leftrightarrow & \bmzeta_{L+\ell} \in [0,1]^p.
\eqq
Combining these expressions proves the assertion. \qed

To fully justify the use of the interior-point method of Portnoy and Koenker (1997), we 
need to verify that the primal problem  (\ref{primal})  can also be put into the required form. 
This is confirmed by the following theorem.

\begin{thm}
The primal problem (\ref{primal}) can be rewritten as
\eqn
\min\{ \ba^T \bmth | \bD \bmth + \bz - \bw = \bb; \bmth \in \bbR^{pK}; \bz,\bw \in \bbR_+^{nL+pL}\},
\label{primal2}
\eqqn
where $\bmth := \bmgam-\bmdel$, $\bz := [\bu_1^T,\dots,\bu_L^T,2\bs_1^T,\dots,2\bs_L^T]^T$,
and $\bw := [\bv_1^T,\dots,\bv_L^T,2\br_1^T,\dots, 2\br_L^T]^T$.
\end{thm}

\noindent
{\it Proof}. Observe that the equality constraints in (\ref{primal}) can be written as 
\eqn
\bD \bmth + \bz - \bw = \bb.
\label{ec}
\eqqn
Under these constraints, we have $\bv_\ell  = \bX \, \bPhi(\tau_\ell) \bmth + \bu_\ell - \by$ and $ \br_\ell = 
c_\ell \, \ddot{\bPhi}(\tau_\ell) \bmth + \bs_\ell$. Substituting these expressions in (\ref{primal}) yields
\eq
\bc^T \bmxi  & = & \sum_{\ell=1}^L \{ \tau_\ell \1_n^T \bu_\ell + (1-\tau_\ell) \1_n^T \bv_\ell + \1_p^T \br_\ell + \1_p^T \bs_\ell \} \\
&=&  \sum_{\ell=1}^L \{ \tau_\ell \1_n^T \bu_\ell +  (1-\tau_\ell) \1_n^T  ( \bX \, \bPhi(\tau_\ell) \bmth +  \bu_\ell - \by) 
+ \1_p^T (c_\ell \, \ddot{\bPhi}(\tau_\ell) \bmth + \bs_\ell) + \1_p^T \bs_\ell \} \\
& = &  \sum_{\ell=1}^L \{ (1-\tau_\ell) \1_n^T  \bX \, \bPhi(\tau_\ell) \bmth + c_\ell \1_p^T \ddot{\bPhi}(\tau_\ell) \bmth
+ \1_n^T \bu_\ell  + 2 \1_p^T \bs_\ell - (1-\tau_\ell)  \1_n^T  \by \} \\
& = & \ba^T \bmth +  \| \bz \|_1 + \text{constant}.
\eqq
The assertion follows upon noting that $\bz$ is not a free parameter.under the condition (\ref{ec}) 
because it is equivalent to $\bD \bmth - \bw \le \bb$ and $\bz := \bb - (\bD \bmth - \bw)$. \qed.

The primal-dual pair given by (\ref{dual}) and (\ref{primal2}) are in the canonical form required by 
the interior-point algorithm of Portnoy and Koenker (1997). This algorithm solves the primal-dual pair jointly by using Newton's method in which positivity constraints are enforced by log barriers (Koenker and Ng 2005). 
An implementation of this algorithm as a FORTRAN code {\tt rqfnb.f}  is available in the `quanreg' package. 
The {\tt rq.fit.fnb} function in this package invokes the FORTRAN code to compute the solution to 
the ordinary QR problem. For the SQR problem, it suffices to modify  the {\tt rq.fit.fnb} function 
using the properly modified input variables.

 The {\tt rq.fit.fnb} function was developed by Portnoy and Koenker (1997) 
for solving the ordinary QR problem $\min_{\bmbeta} \sum_{t=1}^n \rho_\tau(y_t - \bx_t^T \bmbeta)$,
which has a dual formulation of the form 
\eq
\max \{ \by^T \bmzeta | \bX^T \bmzeta = (1-\tau) \bX^T \1_n; \bmzeta \in [0,1]^n\}.
\eqq
Because the dual problem for SQR is given by (\ref{dual}), it suffices to replace the response vector $\by$
by $\bb$, the design matrix $\bX$ by $\bD$, and the right-hand-side vector $(1-\tau) \bX^T \1_n$ in the equality constraints by $\ba$. In addition, we  set the initial value of  $\bmzeta$ to $[(1-\tau_1) \1_n^T,\dots,(1-\tau_L) \1_n^T,0.5 \1_{pL}^T]^T$. We call the resulting R function {\tt rq.fit.fnb2}. 
For solving the SQR problem, the interior-point 
algorithm  turns out to be much more efficient 
than  standard LP solvers, such as the {\tt lp} function, 
in terms of computer memory and  time. 

To develop a data-driven method for selecting the smoothing parameter $c$, we adopt a technique used by 
Koenker et al.\  (1994)  (see also Koenker 2005, p.\ 234) for nonparametric quantile 
regression at a fixed quantile level.  In this technique, 
the mean objective function of quantile regression at the fitted values
is treated as the fidelity measure, analogous to the standard error of the residuals  in linear regression, and the number of points interpolated (or closely approximated) by the fitted values is treated as the complexity measure, similar to the number of parameters in linear regression. 
We extend the fidelity and complexity measures to the SQR problem. 
Let $v_c(\tau_\ell) := n^{-1} \sum_{t=1}^n \rho_{\tau_\ell} (y_t - \bx_t^T \hat{\bmbeta}(\tau_\ell))$ be 
the fidelity measure at $\tau_\ell$,  and $m_c(\tau_\ell)$, denoting the number of points 
closely approximated by the fitted values $\bx_t^T \hat{\bmbeta}(\tau_\ell)$, be the complexity measure 
at $\tau_\ell$. Then, the resulting
Bayesian information criterion (BIC), also known as Schwarz information criterion (SIC), takes the form
\eqn
{\rm BIC}(c) := 2 n \log\bigg( L^{-1} \sum_{\ell=1}^L v_c(\tau_\ell) \bigg)
+ \log n \, \bigg( L^{-1} \sum_{\ell=1}^L m_c(\tau_\ell) \bigg).
\label{BIC}
\eqqn
By the same token, we obtain the AIC criterion which replaces $\log n$ in (\ref{BIC}) with  2. 
As usual, BIC always imposes a heavier penalty
on the complexity than AIC for sufficiently large $n$ such that $\log n > 2$. As a result, 
BIC tends to produce a smoother estimate than AIC.

For convenience, we reparameterize $c$ by {\tt spar} in a  way similar to the
 smoothing parameter in the R function {\tt smooth.spline} (R Core Team 2024), 
 i.e., $c := r \times 1000^{{\tt spar}-1}$, where  
\eq
r := \frac{n^{-1}  \sum_{\ell=1}^L \| \bX \bPhi(\tau_\ell)\|_1}
{ \sum_{\ell=1}^L w_\ell \|  \ddot{\bPhi}(\tau_\ell) \|_1}.
 \eqq 
 In addition, we  use the {\tt smooth.spline} function to generate the knots from a given set of quantile levels, and use the {\tt splineDesign} function in the R package `splines' (R Core Team 2024) to generate the corresponding cubic spline basis functions and their second derivatives.

\section{Examples}

\begin{figure}[t]
\centering
\includegraphics[height=5in,angle=-90]{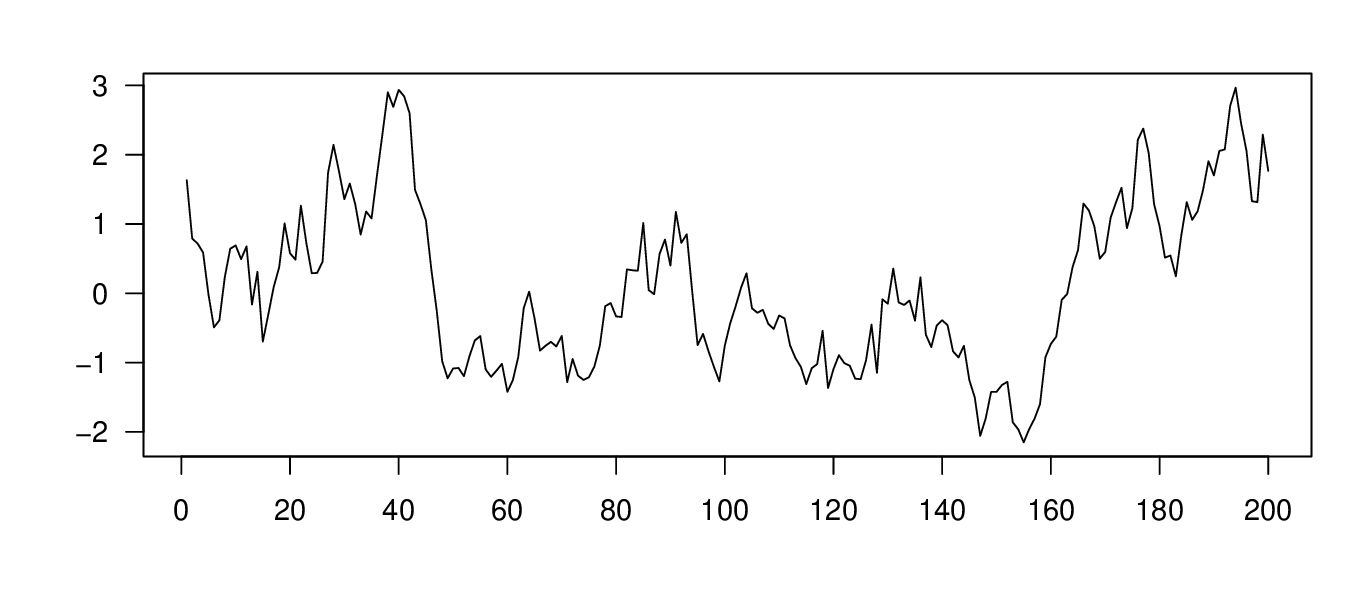} 
\vspace{-0.2in}
\caption{A simulated time series according to  the QAR model (\ref{QAR})  ($n=200$). } \label{fig:QAR}
\end{figure}

First, we evaluate the SQR method by a simulated data set of time series based on a so-called 
quantile autoregressive (QAR) process which satisfies
\eqn
y_t = a_0(U_t) + a_1(U_t) \, y_{t-1}.
\label{QAR}
\eqqn
In this model, $\{ U_t \}$ is a sequence of i.i.d.\ $U(0,1)$ random variables, $a_0(u)$ takes the form 
of the quantile function of $N(0,0.4^2)$, $a_1(u)$ is a piecewise-linear function 
$0.85 + 0.1u + 0.25 (u-0.5) {\cal I}(u > 0.5)$.
This QAR model is a slight modification of the example in Koenker (2005, p.\ 262) where 
$a_0(u)$ is the quantile function of $N(0,1)$ and $a_1(u)$ is a linear function $0.85 + 0.25 u$.
The conditional quantile function of $y_t$ given $y_{t-1}$ can be written as
\eq
Q_{y_t}(\tau \mid y_{t-1} ) = a_0(\tau) + a_1(\tau) \, y_{t-1}, \quad 0 < \tau < 1.
\eqq
Figure~\ref{fig:QAR} shows a series simulated from this QAR model ($n=200)$.

\begin{figure}[p]
\centering
\includegraphics[height=5in,angle=-90]{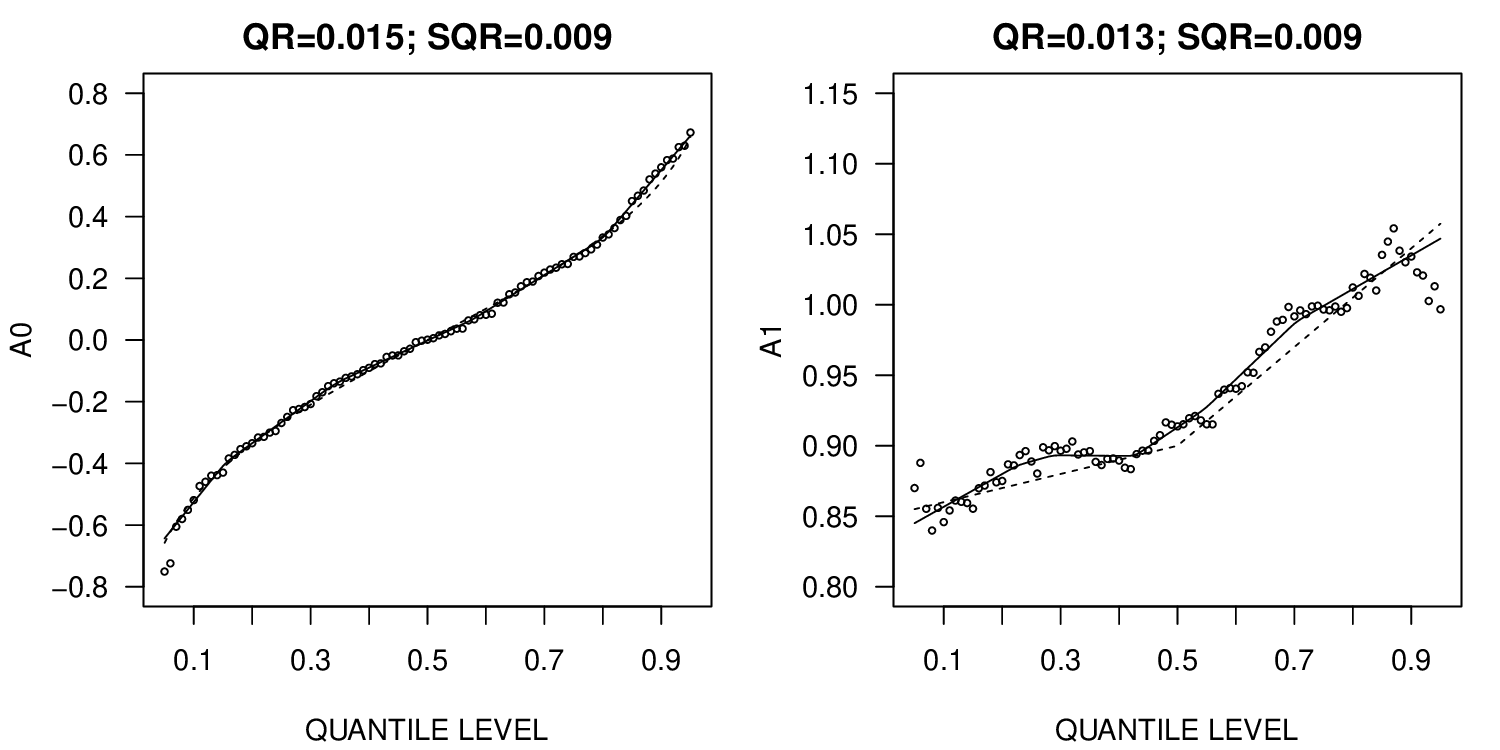} 
\centerline{(a)\hspace{2.25in}(b)}
\includegraphics[height=5in,angle=-90]{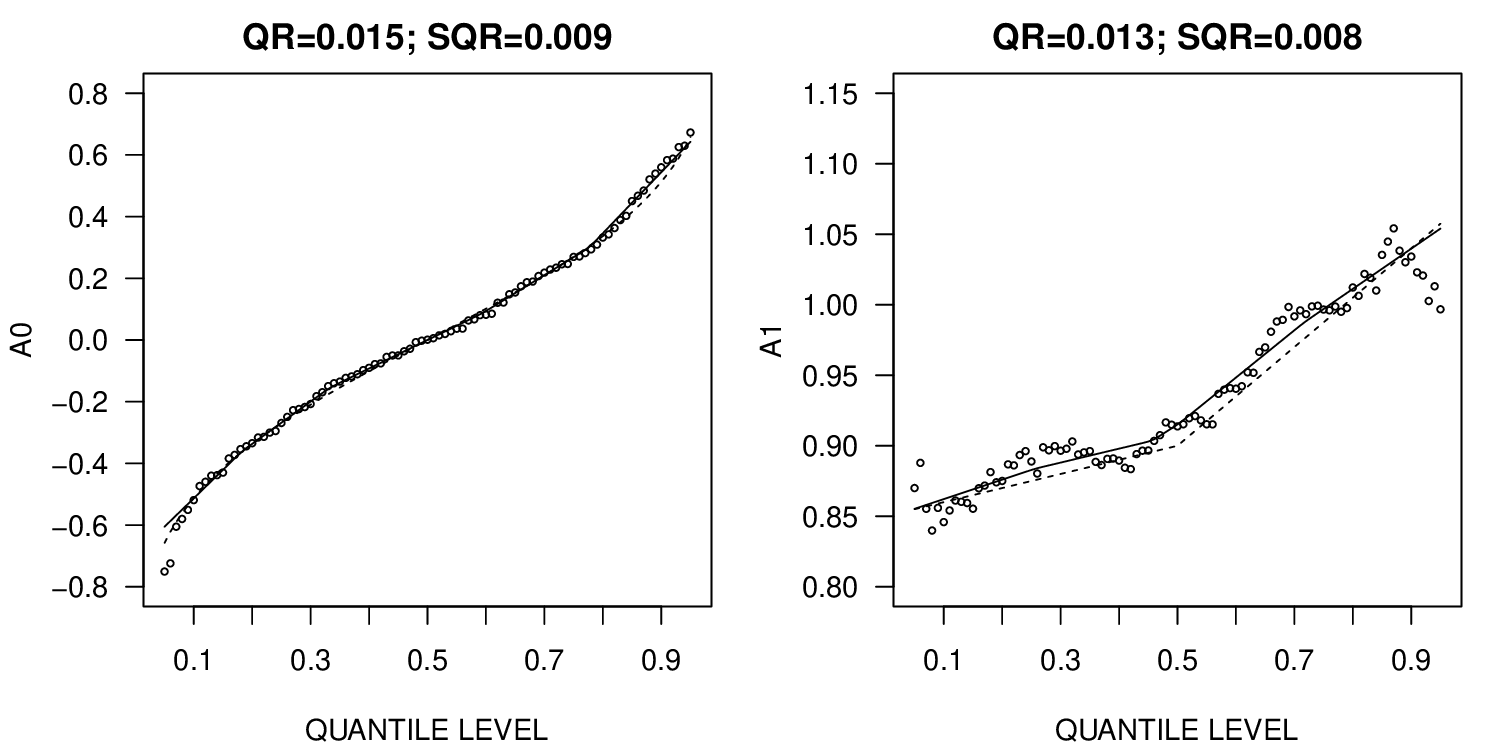} 
\centerline{(c)\hspace{2.25in}(d)}
\caption{SQR estimates of $a_0(\cdot)$ and $a_1(\cdot)$ in the QAR model (\ref{QAR}) 
from the time series shown in Figure~\ref{fig:QAR}.  The smoothing parameter is chosen by (a)-(b)  AIC and (c)-(d) BIC. Open circles depict the QR estimates. Dashed lines depict the true values.  } \label{fig:QAR:coef}
\end{figure}

Figure~\ref{fig:QAR:coef} depicts the SQR estimates and the QR estimates of $a_0(\cdot)$ and $a_1(\cdot)$ 
from this series on the quantile grid $\{ 0.05,0.06,\dots,0.95\}$. For SQR, the smoothing 
parameter is selected by AIC and BIC.  We employ the absolute errors, $L^{-1} \sum_{\ell=1}^L |\hat{a}_j(\tau_\ell) - a_j(\tau_\ell)|$ $(j=0,1)$, to measure the accuracy of these estimates.
For the SQR estimates with AIC, the errors are both 0.009; for the SQR estimates with BIC, 
the errors are 0.009 and 0.008, respectively. The corresponding errors 
of the QR estimates are 0.015 and 0.013.  Visual inspection of Figures~\ref{fig:QAR:coef}(b) and (c)
shows that the SQR with BIC follows the shape of $a_1(\cdot)$ more faithfully than the SQR with AIC.

\begin{figure}[t]
\centering
\includegraphics[height=2.5in,angle=-90]{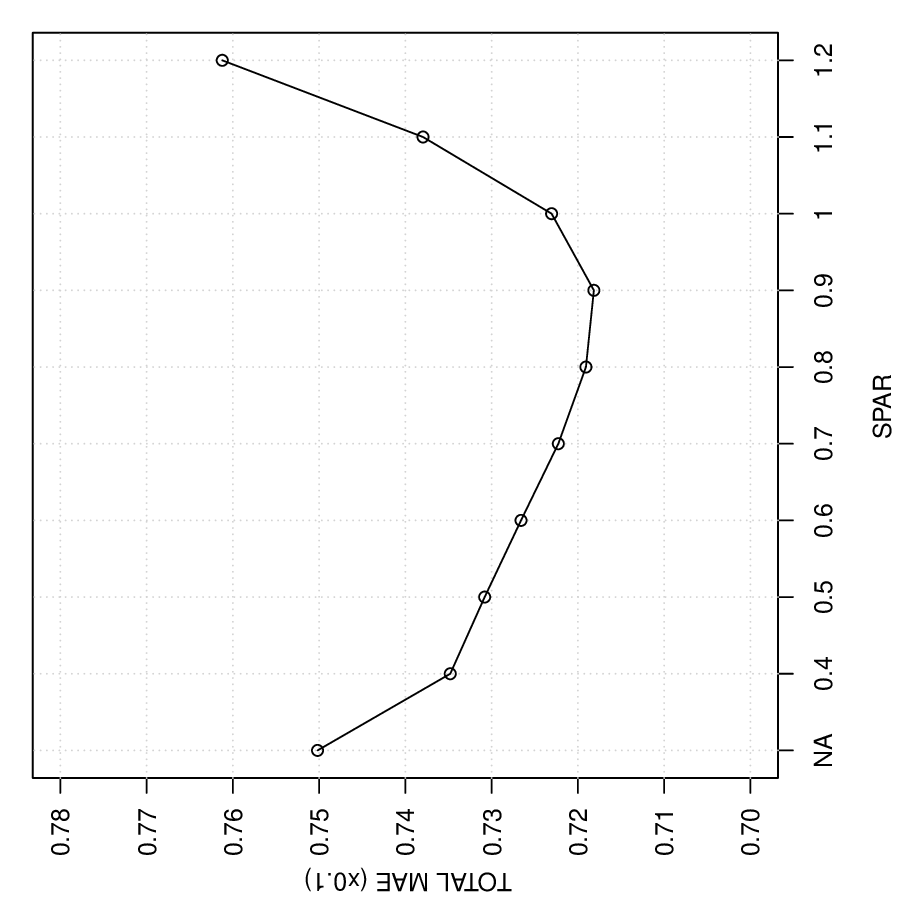} 
\caption{Total mean absolute error for estimating the functional coefficients  in the QAR model (\ref{QAR}) by SQR with different values of smoothing parameter {\tt spar}. NA stands for the ordinary QR. Results are based on 1000 Monte Carlo runs with $n=200$. } \label{fig:QAR:err}
\end{figure}

A more comprehensive analysis of the estimation accuracy is presented in Figure~\ref{fig:QAR:err}.
Based on 1000 Monte Carlo runs, Figure~\ref{fig:QAR:err} shows the total mean 
absolute error $\text{MAE}_0 + \text{MAE}_1$ against the value of smoothing parameter {\tt spar}, where 
\eq
\text{MAE}_j := E \bigg\{ L^{-1} \sum_{\ell=1}^L ( |\hat{a}_j(\tau_\ell) - a_j(\tau_\ell)| \bigg\} \quad (j=0,1).
\eqq
 As can be seen, the SQR estimates outperform the QR estimates
for a range of smoothing parameter values, with the best choice being {\tt spar} = 0.9.  
Table~\ref{tab:err} further demonstrates that the data-driven criteria AIC and BIC are both able to 
select a smoothing parameter for SQR to outperform QR, with BIC yielding a smaller
total MAE for both sample sizes. The improved accuracy of SQR over QR comes largely from estimating $a_1(\cdot)$.

\begin{table}[t]
\begin{center} 
\caption{Mean Absolute Error of SQR Estimates with AIC and BIC 
for the QAR Model (\ref{QAR}) } \label{tab:err}
{\footnotesize
\begin{tabular}{c|ccc|ccc|ccccccc} \hline
& \multicolumn{3}{c|}{$\text{MAE}_0$ } & \multicolumn{3}{c|}{$\text{MAE}_1$} 
&  \multicolumn{3}{c}{Total $\text{MAE}$} \\
$n$  & QR & SQR-AIC & SQR-BIC & QR &  SQR-AIC & SQR-BIC & QR &  SQR-AIC & SQR-BIC  \\  \hline
 200 & 0.0377 & 0.0367 & 0.0366  & 0.0374  & 0.0356 & 0.0351 &  0.0751 & 0.0723 & 0.0717 \\
 500 & 0.0221 & 0.0217 & 0.0221  &  0.0208 & 0.0198 & 0.0194  & 0.0429 & 0.0415 & 0.0411 \\
\hline
\end{tabular} 
}
\end{center}
{\scriptsize 
\begin{center}
\begin{minipage}{6in}
Results are based on 1000 Monte Carlo run. 
\end{minipage}
\end{center}
}
\end{table}

Having validated the performance of the SQR estimator with simulated data, we then present some  real-data examples in the remainder of this section.

The first real-data example is the Engel food expenditure data that comes with the `quantreg' package (Koenker 2005, pp.\ 300--302). It is assume that 
the household expenditure on food, $y$, obeys a quantile regression model  
\eq
Q_{y}(\tau \mid x) = \beta_1(\tau) + \beta_2 (\tau) \,  (x - \mu),
\eqq
where  $x$ is the household income ($\times 1000$) with mean  $\mu$. The data set contains 
$n=235$ records.

\begin{figure}[t]
\centering
\includegraphics[height=5in,angle=-90]{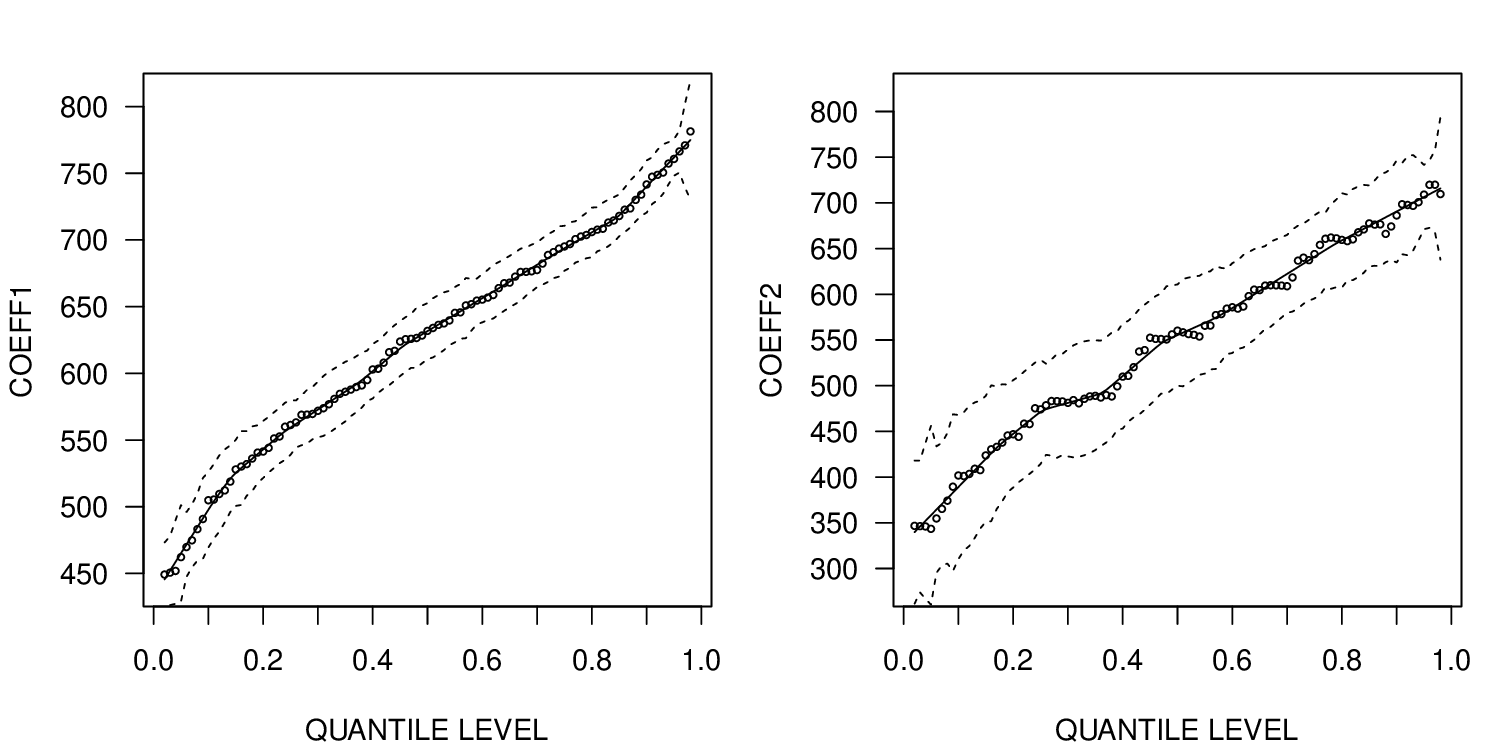} \\
\includegraphics[height=5in,angle=-90]{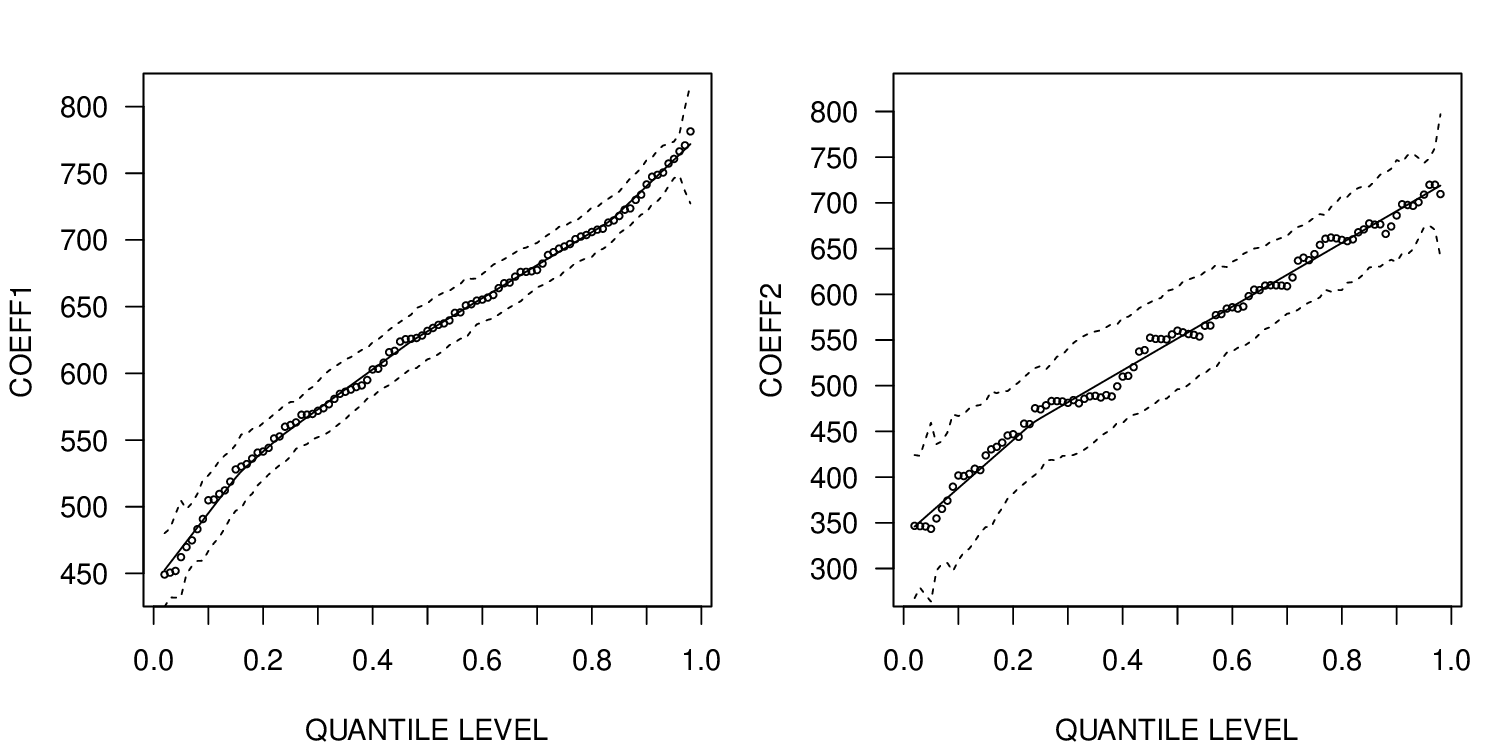}
\caption{SQR estimates of the functional coefficients $\beta_1(\cdot)$ (left) and $\beta_2(\cdot)$ (right) 
for the  Engel food expenditure data. Smoothing parameters are selected by AIC (top) 
and BIC (bottom). Dashed lines depict a 95\% pointwise confidence band. 
Open circles depict the QR estimates.  } \label{fig:engel}
\end{figure}

By following Koenker (2005), we estimate the coefficients $\beta_1(\cdot)$ and $\beta_2(\cdot)$ 
by QR on the quantile grid  $\{ 0.02,0.03,\dots,0.98\}$ with $\mu$ replaced 
by the sample mean of income. These estimates are compared with 
the SQR estimates obtained with automatically selected smoothing parameter by AIC and BIC.
The results are shown in Figure~\ref{fig:engel} as functions of $\tau$.
The confidence band for the SQR estimates is constructed according to the asymptotic normality  of quantile regression estimates 
under the non-iid settings (Koenker 2005, p.\ 34), as implemented by the {\tt summary.rq} function
 in the R package `quantreg'  with the option {\tt se=`nid'}. In comparison with the QR estimates (open circles
in Figure~\ref{fig:engel}), the SQR estimates are less noisy and more  visually appealing. As expected, BIC produces a ``smoother'' estimate than AIC due to a heavier penalty
on the model complexity. It seems particularly appropriate for BIC to smooth out
the wiggly pattern of the QR estimates between $\tau=0.2$ and 0.6, which AIC fails to do.

\begin{figure}[p]
\centering
\includegraphics[height=3in,angle=-90]{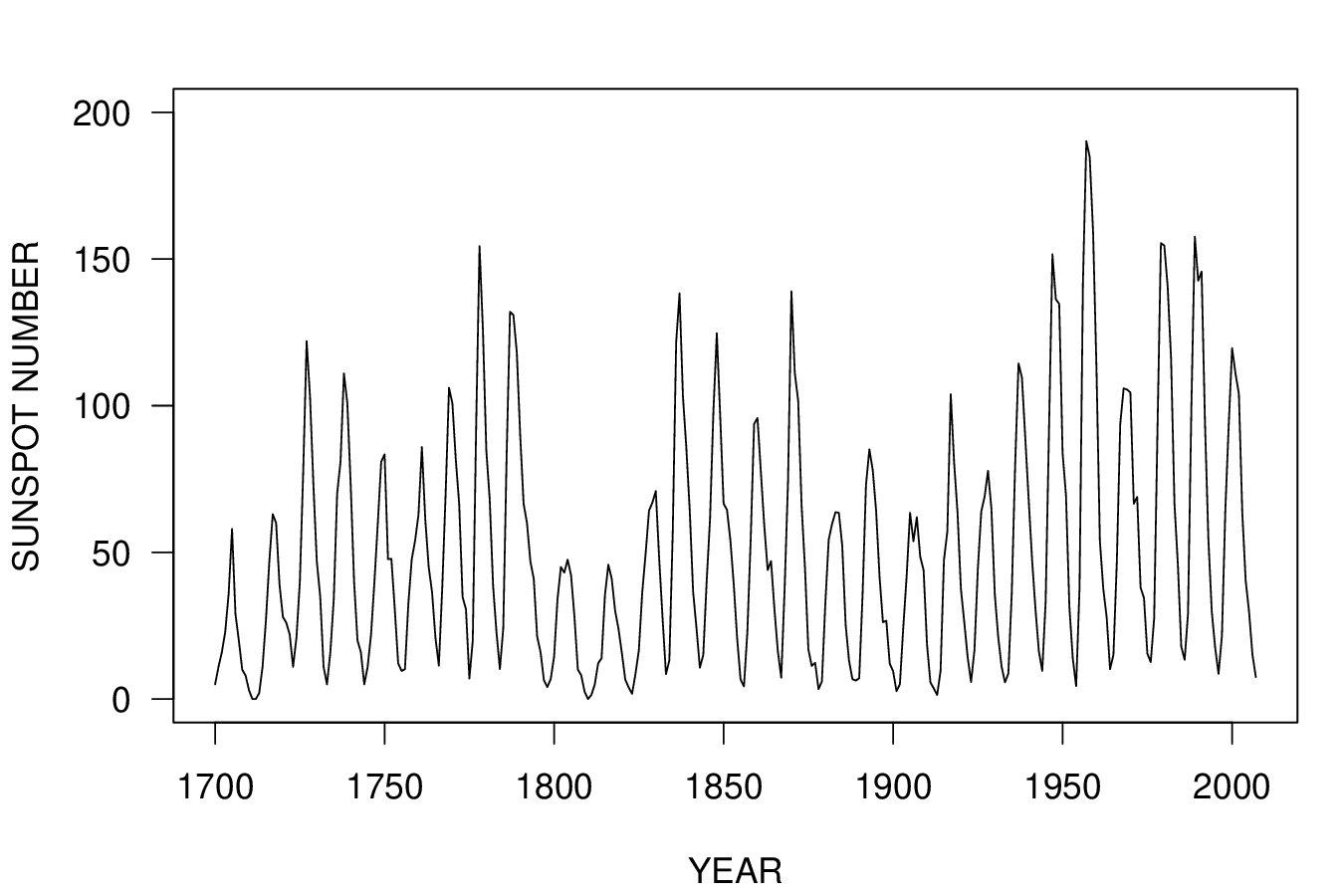} 
\includegraphics[height=3in,angle=-90]{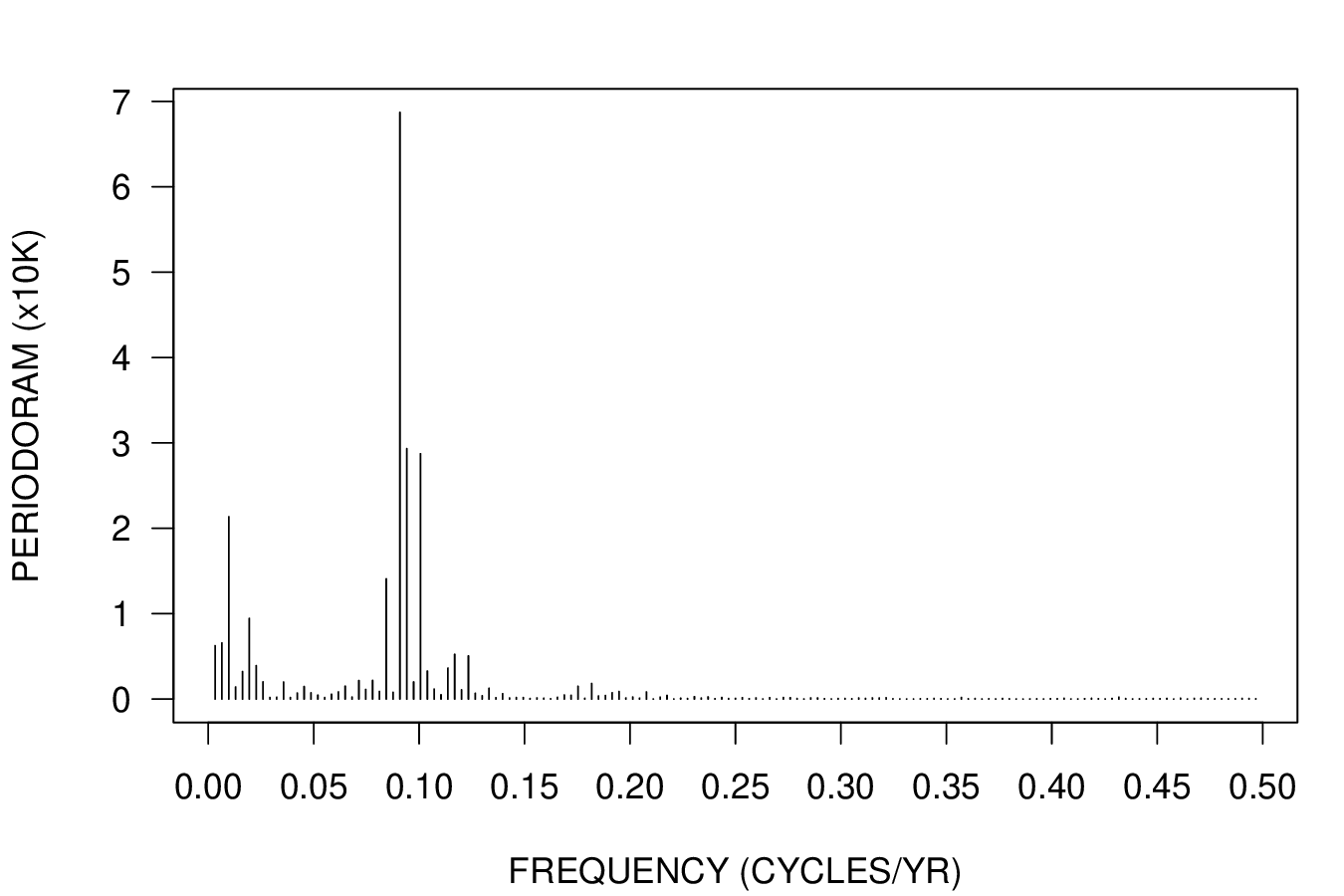} 
\vspace{-0.1in}
\caption{Time series of yearly sunspot numbers from year 1700 to 2007  $(n=308)$ and its conventional periodogram.  } \label{fig:sunspot}
\centering
\includegraphics[height=3in,angle=-90]{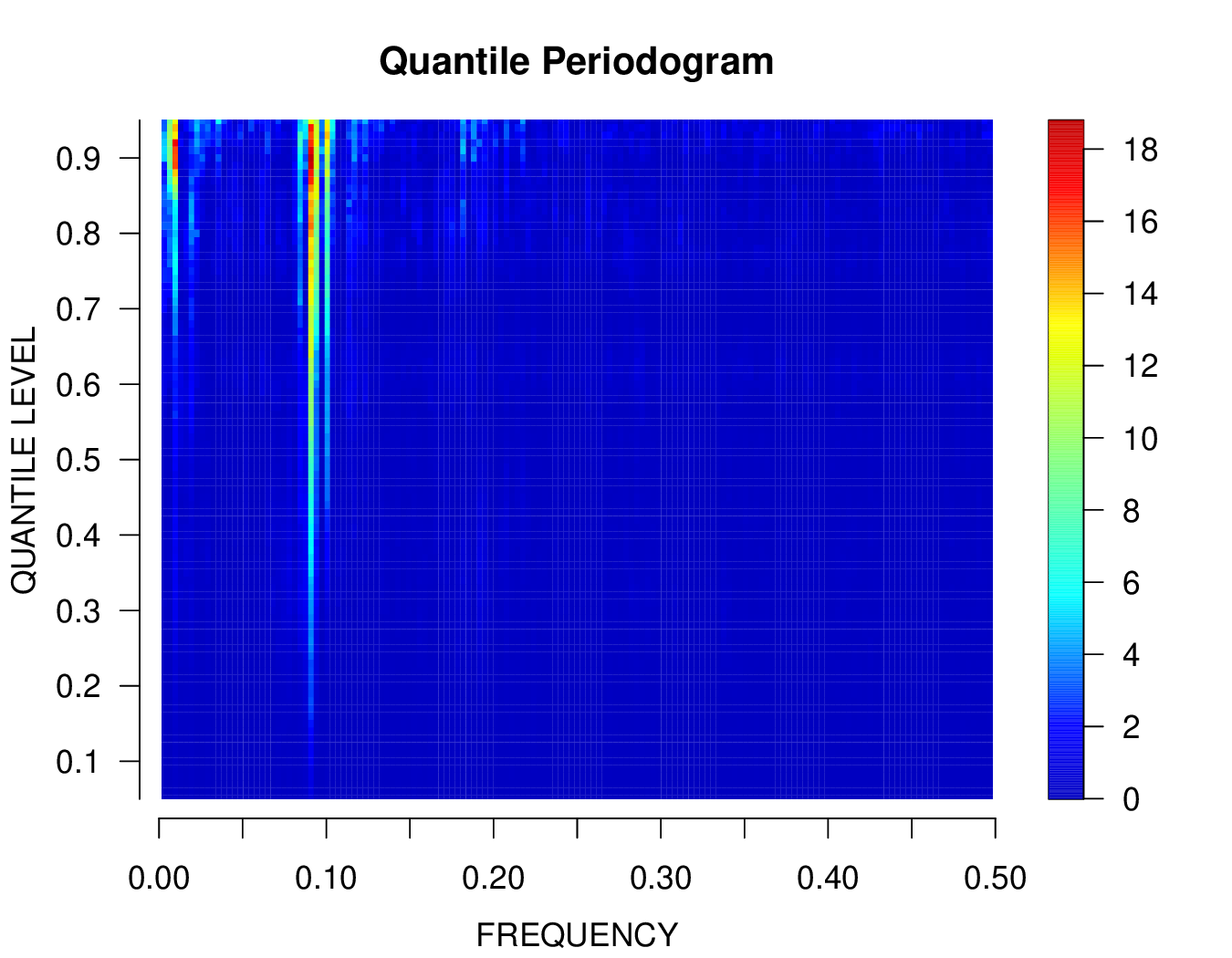} 
\includegraphics[height=3in,angle=-90]{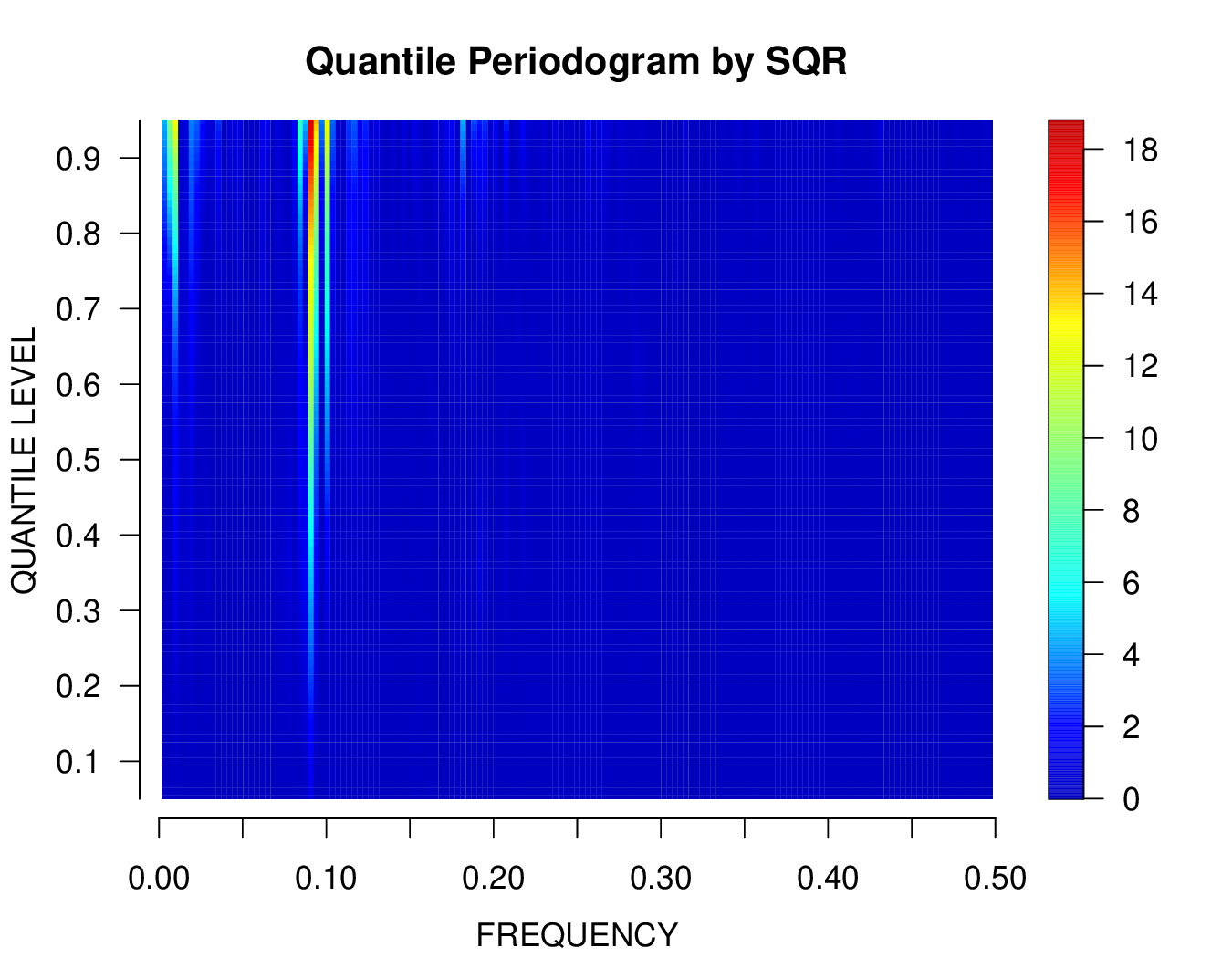} 
\vspace{-0.1in}
\caption{Quantile periodogram of yearly sunspot numbers constructed by trigonometric quantile regression
using QR (left) and SQR with BIC (right). } \label{fig:qper}
\includegraphics[height=3in,angle=-90]{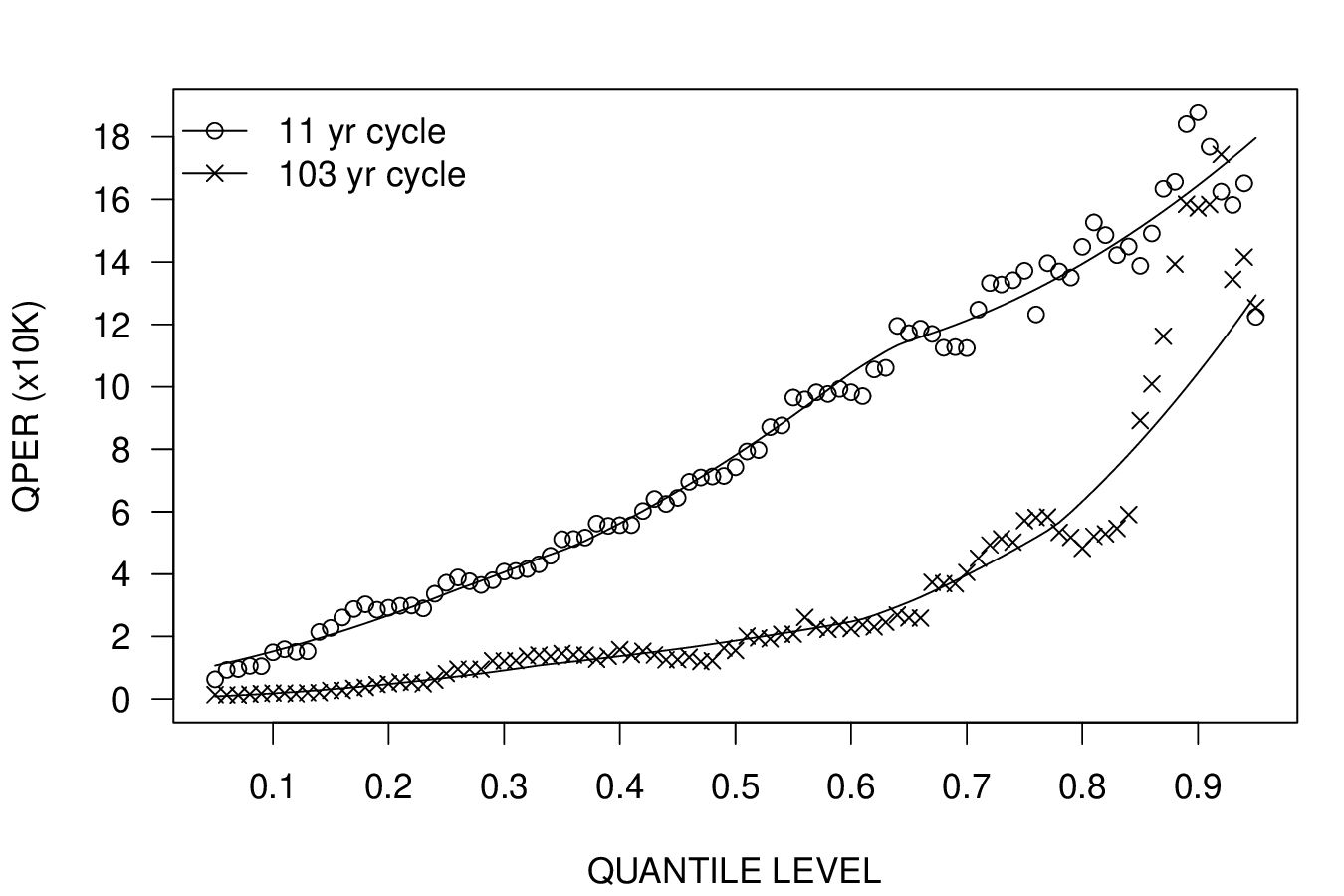} 
\vspace{-0.1in}
\caption{Cross-section of quantile periodogram, constructed by SQR with BIC, of yearly sunspot numbers at 11 year and 103 year frequencies. Open circles  depict the result from QR. } \label{fig:qper2}
\end{figure}

The second real-data example is the time series of yearly sunspot numbers (Figure~\ref{fig:sunspot}).
This series was examined in Li (2012; 2014) for its periodicity through the so-called quantile periodogram
shown in the left panel of Figure~\ref{fig:qper}. This quantile periodogram is an alternative 
to the conventional periodogram for spectral analysis of time sereis. It is
derived from trigonometric quantile regression instead of Fourier transform. Formally, 
with $f$ taking values in the set of normalized Fourier frequencies in interval $(0,0.5)$ and $\tau$ taking values in the interval $(0,1)$, the quantile periodogram  is defined as $(n/4) [\hat{\beta}_2^2(f,\tau) + \hat{\beta}_3^2(f,\tau) ]$, where 
$\hat{\beta}_2(f,\tau) $ and $\hat{\beta}_3(f,\tau) $ are given by QR
using the  regressor $\bx_t(f) := [1,\cos(2\pi f t), \sin(2\pi f t)]^T$, As a bivariate function 
of $f$ and $\tau$,  the quantile periodogram in the left panel of Figure~\ref{fig:qper} suggests that the sunspot
nubmers not only have the well-known  cycle of 11 years ($f = 28/308 \approx  0.091$), but the magnitude of this cycle exhibits a generally increasing trend with the increase of quantile level, i.e., the cycle tends to be stronger at higher quantiles and weaker at lower quantiles.

Instead of QR, the trigonometric quantile regression can be performed by SQR. 
The resulting quantile periodogram is shown in the right panel of Figure~\ref{fig:qper}.
To construct this quantile periodogram, a single smoothing parameter is employed in 
the trigonometric SQR problems for all frequencies. It is chosen by minimizing 
the average BIC across the frequencies. Needless to say, one may also choose to employ different smoothing parameters for different frequencies to achieve greater flexibility but with higher
computational burden and statistical variability.

Compared to the QR-based quantile periodogram, the SQR-based quantile periodogram appears smoother 
across quantiles, not only around the peak frequencies but also in the background. 
The difference between these periodograms can be further appreciated by inspecting 
the cross-section plot in Figure~\ref{fig:qper2}, where the quantile
periodograms are shown as functions of $\tau$ at two peak frequencies corresponding 
to the 11 year cycle $(f=28/308)$ and the 103 year cycle $(f=2/308)$. Compared to the QR estimtes, 
the SQR estimates depict a better-defined less-noisy trend which grows steadily with $\tau$.
The effect of smoothing is most notable at higher quantiles, especially in the curve  
of  the 103 year cycle. 

The final real-data example is the US infant birth weight data from 
the National Bureau of Economic Research (NBER)\footnote{ \url{https://www.nber.org/research/data/vital-statistics-natality-birth-data}} for year 2022.  A similar data set from year 1997 was used in Koenker (2005, p.\ 20) to demonstrate the usefulness of the quantile regression coefficients as functions of $\tau$. 
By following this study, we extract the data records for black or white
mothers between ages 18 and 45. The study in Koenker (2005, p.\ 20)  included 16 explanatory variables for the birth weight (in grams). We simplify the study by considering only 8 explanatory variables: 
infant's sex (0 for Girl, 1 for Boy), mother's race (0 for White, 1 for Black), 
age (18--45), and marital status (0 for Unmarried, 1 for Married), and 
prenatal medical care named {\tt Precare} (0 if the first visit in the first trimester of the pregnancy, 1 if no prenatal visit, 2 if the  first visit in the second trimester,  3 if the first visit in the last trimester). For simplicity, 
mother's education level, smoking status, and weight gain during pregnancy are not used in the model.
As in  Koenker (2005, p.\ 22), we use a quadratic function to represent the effect of age.

\begin{figure}[p]
\centering
\includegraphics[height=5in,angle=-90]{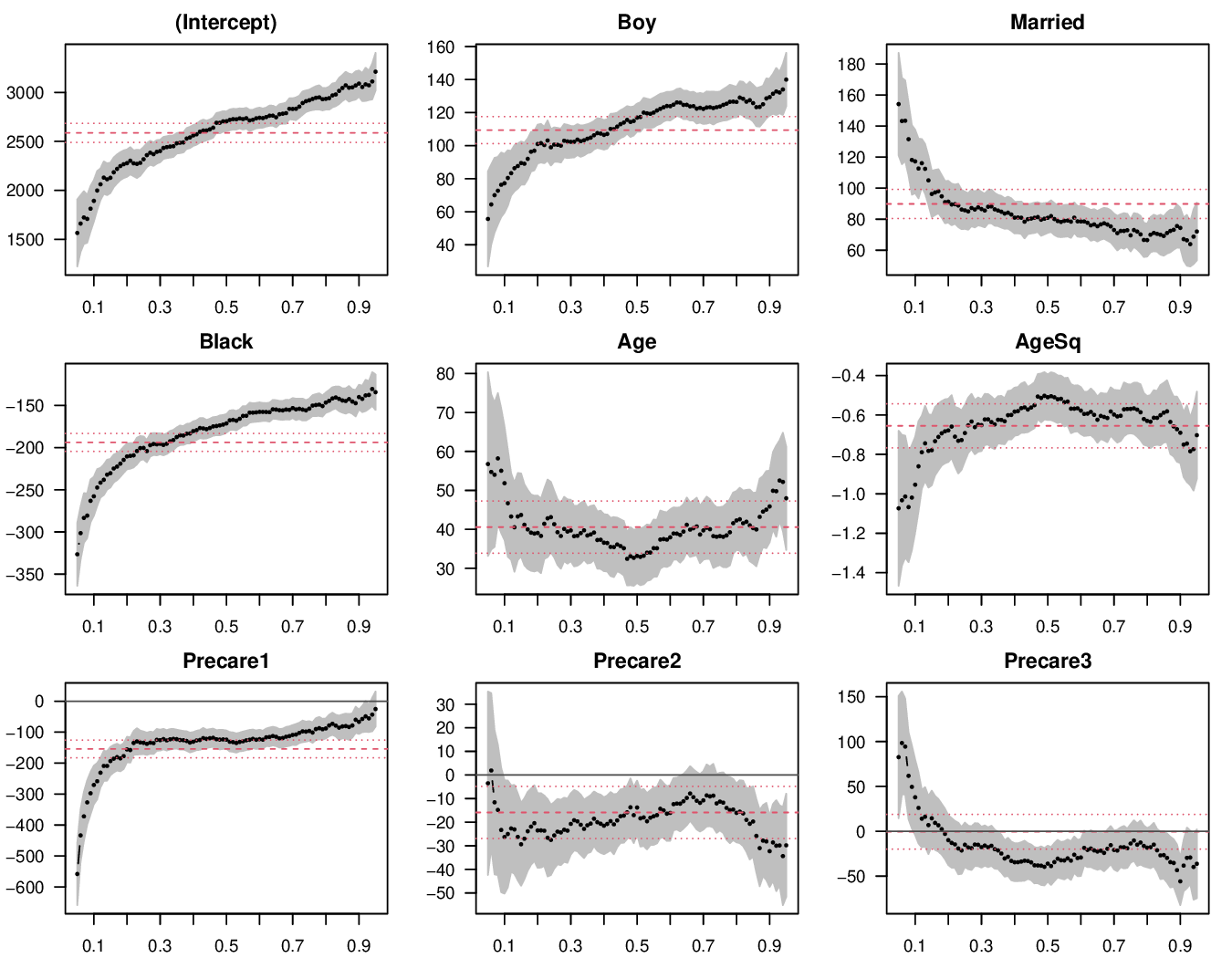} \\
\centerline{(a)}
\vspace{-0.1in}
\includegraphics[height=5in,angle=-90]{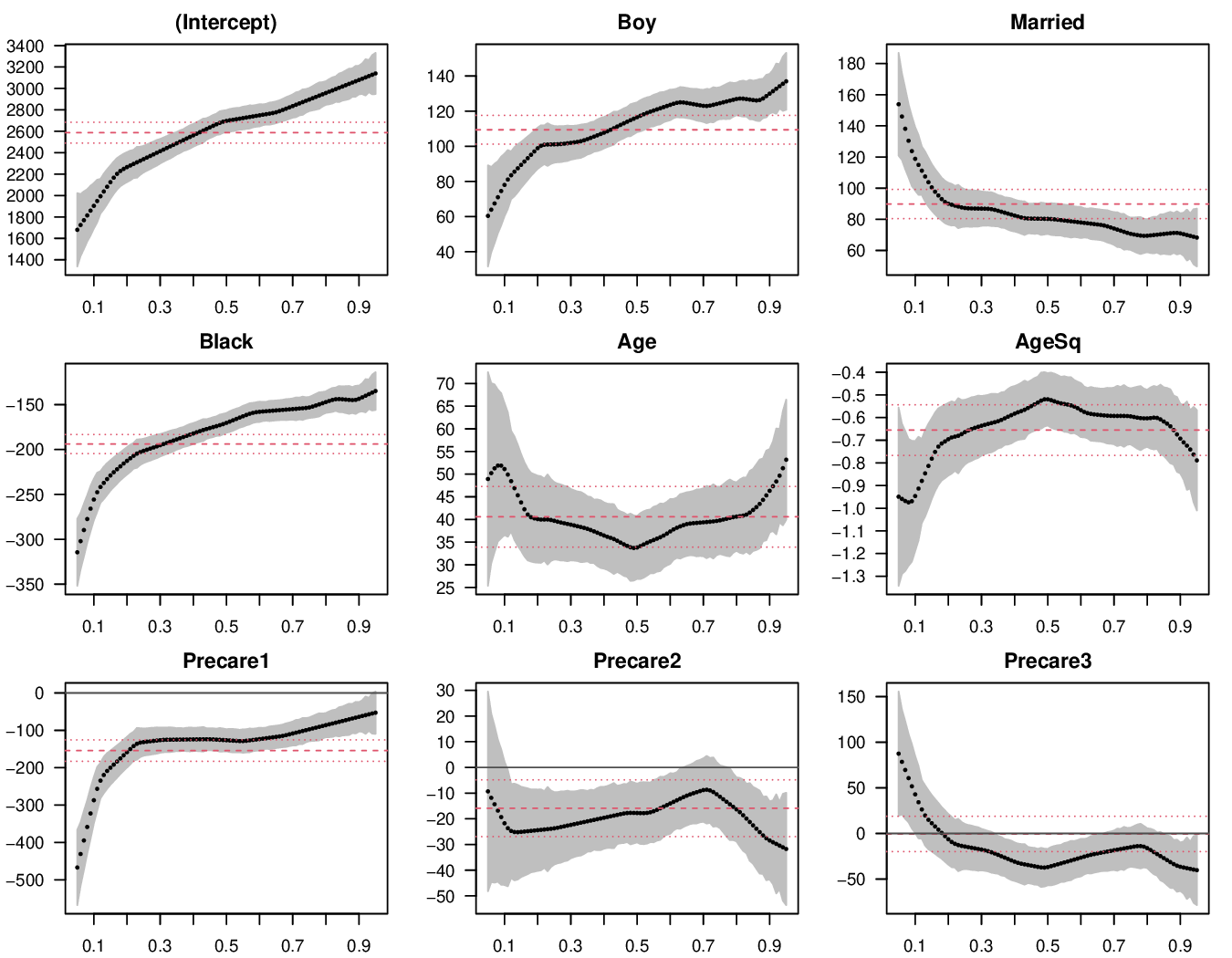} 
\centerline{(b)}
\caption{Quantile regression for birth weight. 
(a) QR estimates. (b) SQR estimates. Shaded gray area depicts a 90\% pointwise confidence band. } 
\label{fig:natality}
\end{figure}

By following Koenker (2005, p.\ 21), we compute the QR estimates on the quantile grid $\{ 0.05,0.06,\dots,0.95\}$. 
These estimates are obtained from a random sample of $n = 50000$ records. 
The corresponding SQR estimates are computed with {\tt spar} $= -1$.
Figure~\ref{fig:natality} shows the QR and SQR estimates as functions of $\tau$,
The 90\%  pointwise confidence bands (shaded grey areas)
for both types of estimates are constructed using the estimated standard errors 
of quantile regression estimates under the iid assumption ({\tt se=`iid'} in {\tt summary.rq}).
As in Koenker (2005, p.\ 21), the dashed and dotted horizontal lines depict the ordinary 
least-squares estimate of the mean effect and a 90\% confidence interval of the estimate. 
The QR estimates in Figure~\ref{fig:natality}(a)  are very similar to those in Koenker (2005, p.\ 21).
The SQR estimates in  Figure~\ref{fig:natality}(b) present the quantile-dependent effect 
of the explanary variables with better-defined and less-noisy patterns.

\section{Gradient Algorithms}

The LP method provides  the exact solution to the SQR problem. However, the artificially inflated number of
decision variables can be a challenge to the computer memory when $n + p$ is large. Indeed, 
the total number of decision variables in the primal-dual pair (\ref{dual}) and (\ref{primal2})  equals 
$2(n+p)L + pK$, which is far greater than the $pK$ variables in the original regression coefficients.
This challenge calls for alternative algorithms that consume less memory but still provide 
reasonably good solutions that may not be exact. 
Gradient algorithms, which directly optimize the objective function in (\ref{sqr2})
with respect to $\bmth$ in $\bbR^{pK}$, are obvious candidates.

Despite the existence of counterexamples  (Asl and Overton 2020), gradient algorithms have been 
successfully used in practice to solve optimization problems with non-smooth objective functions 
(e.g., not everywhere differentiable or continuously differentiable).
This is evidenced, for example, by the effectiveness of such algorithms for training neural network models
involving non-smooth activation functions (Goodfellow et al.\ 2016; Ruder 2016). 
We consider three such algorithms for the SQR problem in (\ref{sqr2}), where the objective function is convex, continuous, and piecewise linear, but not everywhere differentiable.

The first  algorithm is the well-known Broyden–Fletcher–Goldfarb–Shanno
(BFGS) algorithm (Nocedal and Wright 2006).
To supply this algorithm with a gradient function, we set the derivatives of $\rho_\tau(y)$ and $|y|$ 
to zero  when $y=0$. This operation should not have a significant effect on the computed optimization solution 
when the overall gradient is determined jointly by many contributors, most 
of which having properly defined derivatives, as is likely the case for the SQR problem. 
The BFGS algorithm has been found effective for non-smooth problems in practice, but 
a general theory of convergence remains lacking. A mathematical analysis of its behavior for certain
non-smooth functions including $|y|$ can be found in Lewis and Overton (2013).  

The BFGS algorithm is a quasi-Newton method that uses an approximate 
Hessian matrix to capture the local curvature of the objective function together with a line search
to find the optimal step size for each iteration.
The {\tt optim} function in the R package `stats' (R Core Team 2024) has an option for the BFGS algorithm. This implementation employs a backtracking line search strategy and 
works remarkably well in our trigonometric SQR experiement.  Another R implementation,  
{\tt bfgs} (\url{https://rdrr.io/rforge/rHanso/man/bfgs.html}), 
takes a more aggressive bisection approach in line search, guided additionally by the so-called 
curvature condition (Nocedal and Wright 2006). Unfortunately, in our trigonometric SQR experiment, 
this implementation tends to terminate prematurely due to failures in line search.

The Hessian matrix update  and the line search in BFGS are computationally expensive.
For  more economical alternatives, we turn to  a basic gradient 
algorithm without the help of Hessian matrix and line search. Due to their computational simplicity,
such algorithms have become increasingly popular for training neural network models 
(Goodfellow et al.\ 2016; Ruder 2016). A successful example is known as ADAM (Kingma and Ba 2015). In this algorithm, the usual gradient in a gradient descent iteration is replaced by an exponentially 
weighted average of all past gradients and normalized  by the square root of an exponentially 
weighted average of squared gradients. The ADAM algorithm may oscillate around the minimizer indefinitely 
because the step size does not go to zero with the iteration. Therefore, it needs to be terminated 
after a predefined number of iterations.

\begin{table}[t]
{\footnotesize
\begin{center} 
\caption{Limited Line Search in the GRAD Algorithm}
\label{tab:grad}
\begin{tabular}{l} \hline
Limited line search, performed periodically in GRAD after a warm-up phase. \\ 
The criterion for accepting a trial step size and the discount factor $b := 0.2$ are the same as 
in {\tt optim} for BFGS. \\
 (i) (ii) $s_0 :=$ default step size,  (iii) (iv) $s_0 :=$ current step size; $\kappa_0 :=$ number of trials. \\ \hline
 $s \leftarrow \min\{ 1, s_0 \times b^{-\lfloor \kappa_0/2 \rfloor} \}$ (initialize trial step size); 
 $\kappa \leftarrow 0$ (initialize trial count) \\
{\tt While} $s$ not accepted and $\kappa < \kappa_0$ {\tt Do} \\
\hspace{0.2in}  $s \leftarrow s \times b$;  $\kappa \leftarrow \kappa + 1$  \\ 
{\tt End While} \\
{\tt Return} (i) (iv) $s$ if accepted $s_0$ otherwise,  (ii) (iii) $s$ if accepted $s_0 \times b$ otherwise\\ 
\hline
\end{tabular} 
\end{center}
}
\end{table}

The third algorithm, which we call GRAD, is an enhanced version of ADAM. In this algorithm, we modify 
ADAM by usng a limited line search to adjust the step size during iteration rather than holding
the step size constant. The line search,  shown in Table~\ref{tab:grad}, follows the same backtracking strategy as in {\tt optim}. However, it is performed not in every iteration but with a much lower frequency and only after a warm-up period; it is also performed with a small number of trials
including both increase and decrease in step size. Furthermore.
there are four options with difference choices for the starting step size 
and the returning step size when none of the trial step sizes are acceptable.
Options (i) and (ii) start with the default step size used in the warm-up period, 
whereas options (iii) and (iv) start with the current step size. When the
line search fails to produce an acceptable step size, options (i) and (iv) fall back
to the default step size, whereas options (ii) and (iii) fall back to the discounted current step size.
Note that only option (iii) offers gradually decreased step sizes when increases are unacceptable
as suggested by the convergence theory of subgradient methods 
for nonsmooth problems (Polyak 1987). None of these options guarantees a monotone reduction in the objective function. Option (i) deviates the least from ADAM, as the step size remains at the default value except when 
a different trial step size becomes acceptable. Option (iv) can produce the same result as option (i)
when a step size different from the default value is never found acceptable.

\section{Experimentation with Gradient Algorithms}

In this section, we conduct two experiments to evaluate the accuracy of the gradient 
algorithms discussed in the previous section as approximations to the LP solution. 

The first experiment uses the Engel food expenditure data discussed earlier. 
Table~\ref{tab:err:grad} contains the total mean absolute errors of the gradient algorithms 
for approximating the two functional coefficients obtained by LP on the quantile grid 
$\{ 0.02,0.03,\dots,0.98\}$. The best performance is achieved by BFGS, which  approximates 
the LP solution closely with 300 iterations (further iteration makes no significant gains). 
ADAM and GRAD are unable to attain such accuracy despite a large number 
of additional iterations. GRAD is an improvement over ADAM in reducing the approximation error. 
Trading memory requirement with computer time is  a typical property
of gradient algorithms when they work. This is the case for BFGS.
 An informal test shows that BFGS takes 50 seconds to 
complete the 300 iterations versus 12 seconds by LP.

\begin{table}[t]
{\small
\begin{center} 
\caption{Approximation Error for  Engel's Food Expenditure Data}
\label{tab:err:grad}
\begin{tabular}{c|cccccccccccccc} \hline
 & \multicolumn{8}{c}{Number of Iterations} \\
Algorithm  & 0 & 50 & 100 & 150 & 200 & 300 & 400 &  500  & 1000  \\  \hline
 BFGS        &   6.9064 & 6.7616 & 6.6951 & 1.0965 & 0.2112 & 0.1316  \\
 ADAM      &   6.9064 & 6.7194 & 6.6910 & 6.6903 & 6.6904 &  6.6904 & 6.6904 & 6.6904  & 6.6904  \\
 GRAD      &   6.9064 & 6.7194 &  6.6835 & 6.6771 & 6.6774 & 6.6027 & 6.6034 & 6.5307 & 6.4270 \\
\hline
\end{tabular} 
\end{center}
}
{\scriptsize 
\begin{center}
\begin{minipage}{5.8in}
QR is used as initial value in iteration 0. For ADAM and GRAD, warm-up phase = 70, 
frequency of step size update = 20, initial step size $s_0 = 0.4$, and discount factor $b = 0.2$.
GRAD uses  option (i) for line search with $\kappa_0 = 5$.
\end{minipage}
\end{center}
}
\end{table}

The second experiment  employs a simulated time series  of length $n=512$ Motivated by the quantile periodogram (Li 2012), 
we compute the regression coefficients $\hat{\beta}_1(\tau,\om),\hat{\beta}_2(\tau,\om)$, and 
$\hat{\beta}_3(\tau,\om)$
on the quantile grid $\tau \in \{0.10,0.11,\dots,0.90\}$ using the trigonometric
regressor $\bx_t(\om)  := [1,\cos(\om t), \sin(\om t)]^T$ at 255 Fourier 
frequencies $\om \in \{ 2\pi v /n: v=1,\dots, \lfloor (n-1)/2 \rfloor \}$. 
The coefficients $\hat{\beta}_2(\tau,\om)$ and $\hat{\beta}_3(\tau,\om)$, which define the quantile periodogram, 
are compared with the corresponding LP solutions.
The average sum of squared errors across the quantiles and frequencies 
is used to measure the accuracy of the gradient algorithm.

\begin{figure}[t]
\centering
\includegraphics[height=3in,angle=-90]{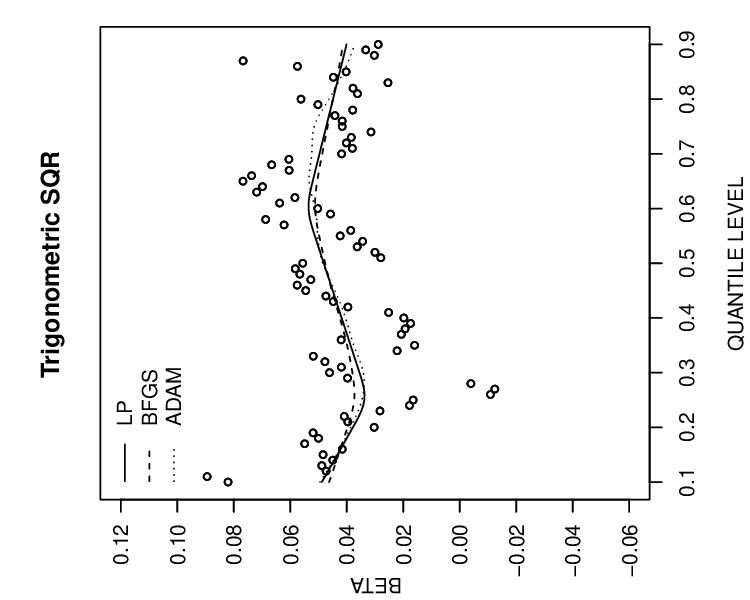}
\includegraphics[height=3in,angle=-90]{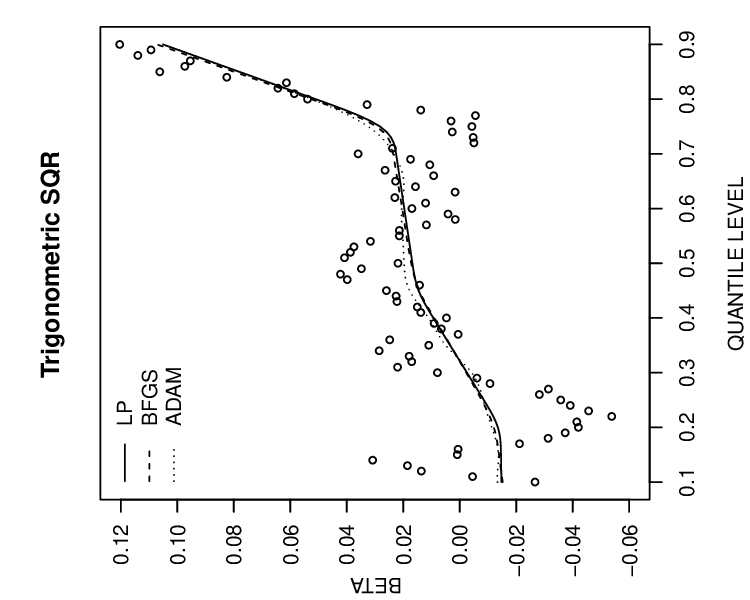}  
\caption{Coefficients of trigonometric SQR at frequency $\om = 2\pi \times 102/512$ computed by LP (solid line), BFGS (dashed line), and ADAM (dotted line). Open circles depict  the QR solution. } \label{fig:sqr}
\end{figure}

The time series $\{ y_t \}$  is generated by a nonlinear mixture model in Li (2020). For brevity, the exact form
of this model is omitted here because it is not important for our discussion. It suffices to say that
the 255 LP solutions have a variety of patterns based on which
we evaluate the approximations produced by the gradient algorithms. 
One of the solutions is shown in Figure~\ref{fig:sqr} together with the approximations produced by
BFGS and ADAM and the corresponding QR solution. Both BFGS and ADAM 
are terminated after 100 iterations. As can be seen, the estimates from BFGS and ADAM 
appear remarkably similar to the estimates from LP.

\begin{figure}[t]
\centering
\includegraphics[height=3in,angle=-90]{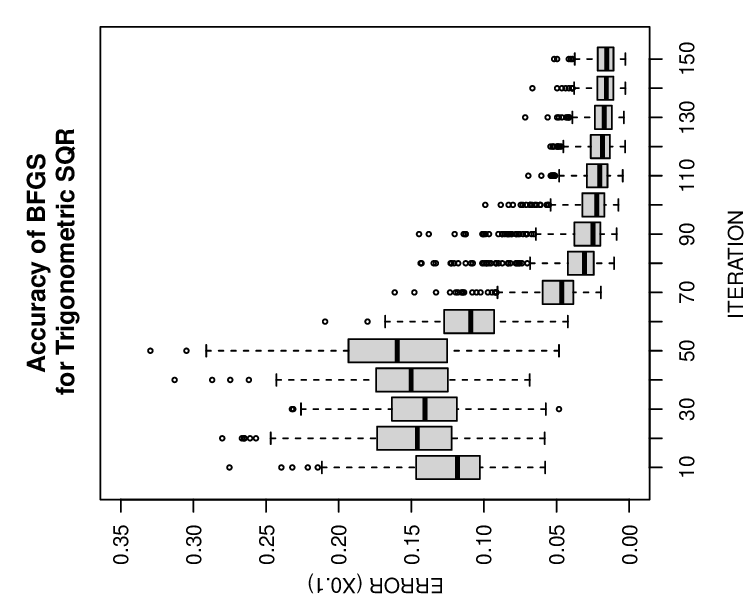}
\includegraphics[height=3in,angle=-90]{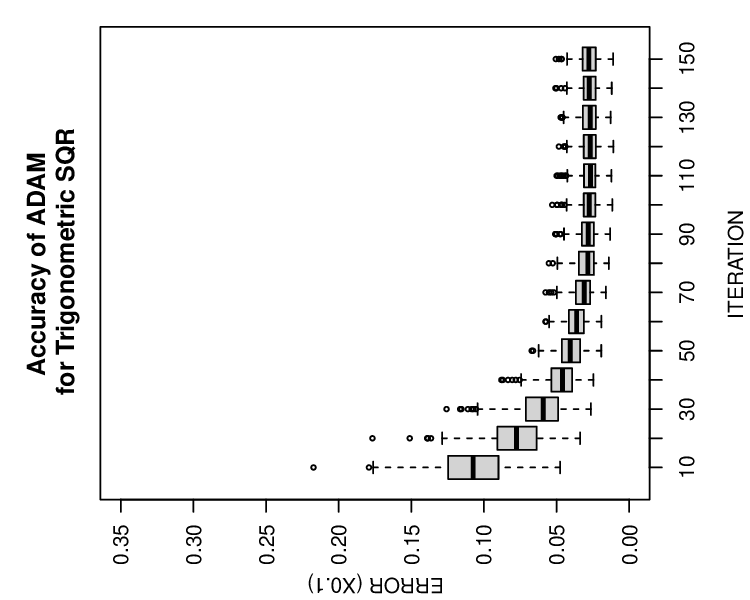}
\caption{Boxplot of approximation errors of trigonometric SQR at 255 frequencies versus the number of iterations computed by BFGS (left) and  ADAM (right). } \label{fig:approx}
\end{figure}

Figure~\ref{fig:approx} shows the the boxplot of approximation errors of BFGS and ADAM 
for all 255 frequencies against the number of iterations. 
This result confirms that both BFGS and ADAM,  with sufficiently large number of iterations, 
provide reasonably good approximations to the LP solution. The BFGS algorithm is able to offer 
a higher accuracy of approximation than ADAM after an initial warm-up period. 
The ADAM algorithm has a more successful start in early stages of the iteration, but stops improving 
beyond a certain number of iterations.

\begin{figure}[t]
\centering
\includegraphics[height=3in,angle=-90]{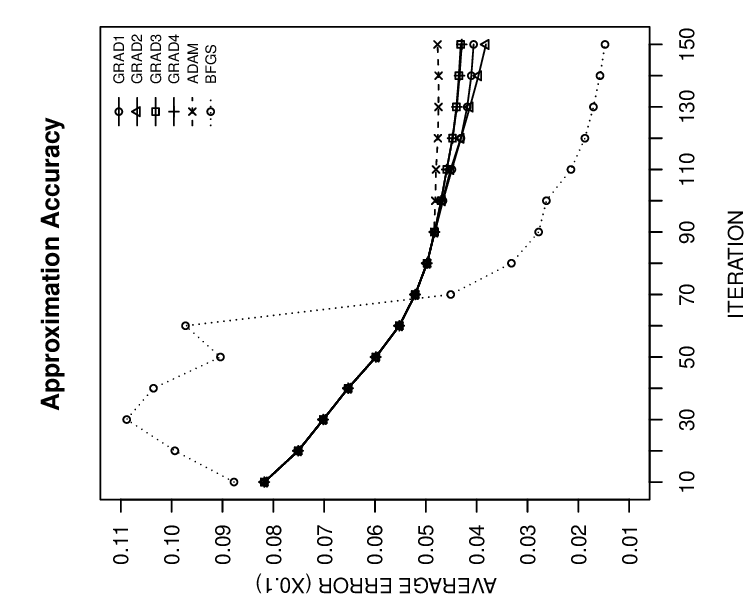}
\includegraphics[height=3in,angle=-90]{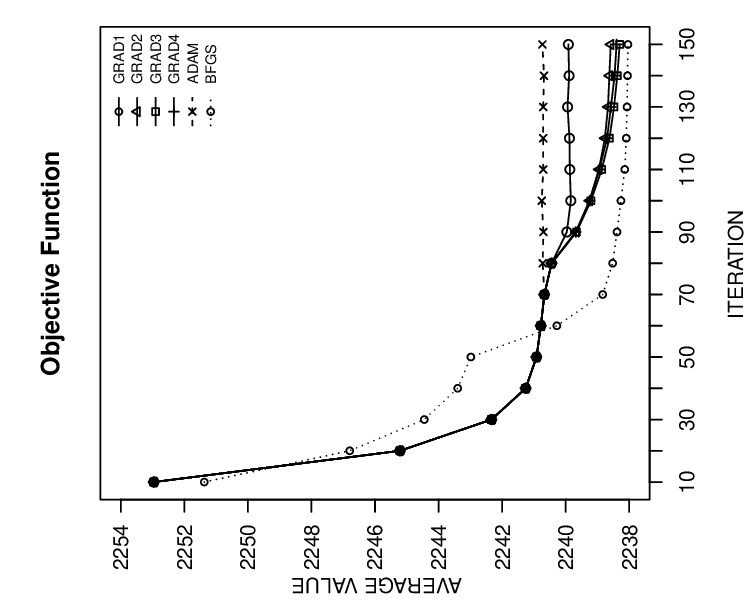}  
\caption{Average approximation error (left) and average objective function (right) 
of trigonometric SQR across 255  frequencies  versus the number of iterations computed by BFGS, 
ADAM, and  GRAD with 4 options. } 
\label{fig:approx3}
\end{figure}

Figure~\ref{fig:approx3} compares GRAD with ADAM and BFGS based on the average approximation error
across all 255 frequencies. Line search in GRAD is performed once every 
10 iterations with 5 trial step sizes after 70 warm-up iterations. 
Thanks to these modifications, GRAD is able to outperform 
ADAM in terms of the approximation error and the objective function after the warm-up iterations.
Among the four options in GRAD, option (iii) is most effective in reducing the objective function, 
whereas option (i) is least effective because it does not shrink the step size 
when the limited line search fails. In terms of reducing the approximation error, 
option (ii) turns out to be most effective.

 It is interesting to observe that after 150 iterations the average approximation error 
 of GRAD remains much larger than that of BFGS,  although the average objective function of GRAD 
 becomes near that of BFGS. Such discrepancies are often observed in problems
where the objective function has a shallow minimum in some  variables. 
With BFGS, possible differences in curvature among variables are equalized by the Hessian matrix.
This enables BFGS to move quickly toward the solution with the same step size for all variables. 
Being a first-order algorithm, neither GRAD nor ADAM possesses this capability. Slow improvement (if any) 
is expected for these algorithms due to the requirement of small step sizes.

\section{Concluding Remarks}

In summary, we consider the problem of fitting linear models by quantile regression at multiple quantile levels where the coefficients of regressors are represented by spline functions of the quantile level and penalized to ensure smoothness across quantiles. Using the $L_1$-norm of second derivatives as the penalty term, the resulting spline quantile regression (SQR) problem can be reformulated and solved as a linear program (LP) by an 
interior-point algorithm. The SQR solution complements the ordinary quantile regression (QR) solution obtained
independently for each quantile level. Our experiments show that an improved estimate over QR can be obtained by SQR when the underlying functional coefficients are suitably smooth.

We also consider three gradient algorithms, BFGS, ADAM, and GRAD, to provide approximations
to the LP solution with less computer memory, Our  experiments show that  it is possible for 
these gradient algorithms, especially BFGS, to produce reasonably good approximations
to the LP solution, but a large number of iterations may be required, especially for ADAM 
and GRAD. The variable step size in GRAD is more effective than the fixed step size
in ADAM. How to improve ADAM and GRAD to achieve a similar performance to BFGS without 
a significant increase of computational burden remains an interesting
problem for future research.

\section*{References}

{\footnotesize
\begin{description} 
\item
Andriyana, Y., Gijbels, I., and Verhasselt, A. (2014). P-splines quantile regression estimation in varying
coefficient models. {\it Test}, 23, 153--194.

\item
Asl, A., and Overton, M. (2020).
Analysis of the gradient method with an Armijo–Wolfe line search on a class of non-smooth convex functions.
{\it Optimization Methods and Software}, 35, 223--242.

\item
Belloni, A., Chernozhukov, V., Chetverikov, D., and Fern\'{a}ndez-Val, I. (2019).
Conditional quantile processes based on series or many regressors.
{\it Journal of Econometrics}, 213, 4--29. 

\item
Berkelaar, M. (2024). Package `lpSolve'.  \url{https://cran.r-project.org/web/packages/lpSolve/lpSolve.pdf}.

\item
Bondell, H., Reich, B. and Wang, H. (2010).
Noncrossing quantile regression curve estimation.
{\it Biometrika}, 97, 825--838.

\item
Goodfellow, I., Bengio, Y., and Courville, A.  (2016).
{\it Deep Learning}. Cambridge, MA: MIT Press.

\item
Hao, M., Lin, Y., Shen, G., and Su, W. (2023). 
Nonparametric inference on smoothed quantile regression process.
{\it Computational Statistics and Data Analysis}, 179, 107645.

\item
He, X. (1997). Quantile curves without crossing. {\it American Statistician}, 51, 186--192.

\item
He, X., Pan, X., Tan, K., and Zhou, W. (2023).
Smoothed quantile regression with large-scale inference. {\it Journal of Econometrics}, 232, 367--388.

\item
Kim, M.-O. (2007). Quantile regression with varying coefficients.
{\it Annals of Statistics}, 35, 92--108.

\item
Kingma, D., and Ba, J. (2015). Adam: a method for stochastic optimization. 
{\it International Conference on Learning Representations}, arXiv:1412.6980.

\item
Koenker, R. (2005). {\it Quantile Regression}. Cambridge University Press, Cambridge.

\item
Koenker, R., and Bassett, G. (1978). Regression quantiles. {\it Econometrica}, 46, 33--50.

\item
Koenker, R., and Ng, P. (2005).
A Frisch-Newton algorithm for sparse quantile regression.
{\it Acta Mathematicae Applicatae Sinica},  21, 225--236.

\item
Koenker, R., Ng, P., and Portnoy, S. (1994). Quantile smoothing splines. {\it Biometrika}, 81, 673--680.

\item
Lewis, A., and Overton, M. (2013). Non-smooth optimization via quasi-Newton methods,
{\it Mathematical Programming}, 141, 135--163.

\item
Li, T.-H. (2012). Quantile periodograms. {\it Journal of the American Statistical Association}, 107, 765--776. 

\item 
Li, T.-H. (2014). {\it Time Series with Mixed Spectra}. Boca Raton, FL: CRC Press.

\item
Li, T.-H. (2020). From zero crossings to quantile-frequency analysis of time
series with an application to nondestructive evaluation. 
{\it Applied Stochastic Models for Business and Industry}, 36, 1111--1130.

\item
Nocedal, J., and Wright, S. (2006). {\it Numerical Optimization}, 2nd edn. New York: Springer.

\item
Oh, H.-S., Lee, T., and Nychka, D. (2011).
Fast nonparametric quantile regression with arbitrary smoothing methods.
{\it Journal of Computational and Graphical Statistics}, 20, 510--526.

\item
Portnoy, S. (1991). Asymptotic behavior of the number of regression quantile breakpoints. 
{\it SIAM journal on scientific and statistical computing}, 12, :867--883.

\item
Portnoy, S., and Koenker, R. (1997). The Gaussian hare and the Laplacian tortoise: computability of squared-error
versus absolute-error estimators. {\it Statistical Science}, 12, 279--300.

\item
Polyak, B. (1987). {\it Introduction to Optimization}, Chapter. 5.  New York: Optimization Software, Inc.

\item
R Core Team (2024). R: A language and environment for statistical
  computing. R Foundation for Statistical Computing, Vienna,
  Austria. \url{https://www.R-project.org/}.

\item
Ruder, S. (2016). An overview of gradient descent optimization algorithms. arXiv:1609.04747.

\item
Wahba, G. (1975). Smoothing noisy data with spline functions.
{\it Numerische Mathematik}, 24, 383--393.

\item
Wu, Y., and Liu, Y. (2009). Stepwise multiple quantile regression estimation
using non-crossing constraints. {\it Statistics and Its Interface}, 2, 299--310.

\end{description}
}

\newpage

\section*{Appendix: R Functions}

The following functions are implemented in the R package `qfa' (version $\ge$ 4.1)
available at \url{https://cran.r-project.org} and \url{https://github.com/thl2019/QFA}.

\begin{itemize}
\item {\tt sqr}: a function that computes the SQR solution on a grid of quantile levels 
by the  interior-point algorithm of Koenker et al.\  (1994) with or without user-supplied smoothing parameter {\tt spar}.
\item {\tt sqdft}: a function that computes the quantile discrete Fourier transform (QDFT) of time series data based on trigonometric SQR solutions on a grid of quantile levels with or without user-supplied smoothing parameter {\tt spar}.
\item {\tt qdft2qper}: a function that converts the QDFT produced by {\tt sqdft} into a quantile periodogram (QPER).
\item {\tt qfa.plot}: a function that produces a quantile-frequency image plot for a  quantile spectrum.
\item {\tt tsqr.fit}: a  low-level function that computes the trigonometric SQR solution on a grid of quamtile levels for a given frequency with a given smoothing parameter {\tt spar}.
\item {\tt sqr.fit.optim}: a function that computes the SQR solution on a grid of quantile levels 
with a given smoothing parameter {\tt spar} by BFGS, ADAM, or GRAD algorithm.
\end{itemize}

\end{document}